\documentclass[11pt,a4paper]{article} 
\usepackage{jcappub} 
\usepackage{amsmath,amsfonts,amssymb}
\usepackage{txfonts}
\usepackage{graphicx}
\usepackage{ifthen}
\usepackage{relsize}
\usepackage{xcolor}

\usepackage{subcaption}
\usepackage[labelsep=period]{caption}
\usepackage[normalem]{ulem} 
\usepackage{hyperref}

\usepackage{mathrsfs}
\usepackage{multirow}
\usepackage{booktabs}

\usepackage{threeparttable}

\makeatletter
\g@addto@macro\bfseries{\boldmath}
\makeatother

\newcommand{\mymatrix}[1]{\mathbf{#1}}

\newcommand{\LB}{\textit{LiteBIRD}}
\newcommand{\lb}{\textit{LiteBIRD}}
\newcommand{\Planck}{\textit{Planck}}
\newcommand{\planck}{\textit{Planck}}
\newcommand{\wmap}{\textit{WMAP}}
\newcommand{\WMAP}{\textit{WMAP}}
\newcommand{\smica}{\texttt{SMICA}}
\newcommand{\nilc}{\texttt{NILC}}
\newcommand{\sevem}{\texttt{SEVEM}}
\newcommand{\commander}{\texttt{Commander}}
\newcommand{\healpix}{\texttt{HEALPix}}
\newcommand{\healpy}{\texttt{healpy}}
\newcommand{\nside}{$N_{\mathrm{side}}$}

\title{\LB\ Science Goals and Forecasts. $E$-mode Anomalies.}

\author[1]{A.\,J.\,Banday,}
\author[2]{C.\,Gimeno-Amo,}
\author[3]{P.\,Diego-Palazuelos,}
\author[4]{E.\,de\,la\,Hoz,}
\author[5,6]{A.\,Gruppuso,}
\author[7]{N.\,Raffuzzi,}
\author[2]{E.\,Martínez-González,}
\author[2]{P.\,Vielva,}
\author[2]{R.\,B.\,Barreiro,}
\author[7,8]{M.\,Bortolami,}
% ...
\affiliation[1]{IRAP, Université de Toulouse, CNRS, CNES, UPS, Toulouse, France}
\affiliation[2]{Instituto de Fisica de Cantabria (IFCA, CSIC-UC), Avenida los Castros SN, 39005, Santander, Spain}
\affiliation[3]{Max Planck Institute for Astrophysics, Karl-Schwarzschild-Str. 1, D-85748 Garching, Germany}
\affiliation[4]{CNRS-UCB International Research Laboratory, Centre Pierre Binétruy, UMI2007, Berkeley, CA 94720, USA}
\affiliation[5]{INAF - OAS Bologna, via Piero Gobetti, 93/3, 40129 Bologna, Italy}
\affiliation[6]{INFN Sezione di Bologna, Viale C. Berti Pichat, 6/2 – 40127 Bologna, Italy}
\affiliation[7]{Dipartimento di Fisica e Scienze della Terra, Università di Ferrara, Via Saragat 1, 44122 Ferrara, Italy}
\affiliation[8]{INFN Sezione di Ferrara, Via Saragat 1, 44122 Ferrara, Italy}
% ...

\abstract{\hfill\newline
Various so-called anomalies have been found in both the \wmap\ and \Planck\ cosmic microwave background (CMB) temperature data that exert a mild tension against the highly successful best-fit 6 parameter cosmological model, potentially providing hints of new physics to be explored. That these are real features on the sky is uncontested.
However, given their modest significance, whether they are indicative of true departures from the standard cosmology or simply statistical excursions, due to a mildly unusual configuration of temperature anisotropies on the sky which we refer to as the \lq\lq fluke hypothesis" cannot be addressed further without new information. 

No theoretical model of primordial perturbations has to date been constructed that can explain all of the temperature anomalies. Therefore, we focus in this paper on testing the fluke hypothesis, based on the partial correlation between the temperature and $E$-mode CMB polarisation signal.
In particular, we compare the properties of specific statistics in polarisation, built from
unconstrained realisations of the $\Lambda$CDM cosmological model as might be observed by the \lb\ satellite, with those determined from
constrained simulations, where the part of the $E$-mode anisotropy correlated with temperature is constrained by observations of the latter. 
Specifically, we use inpainted \Planck\ 2018 \smica\ temperature data to constrain the $E$-mode realisations. 
Subsequent analysis makes use of masks defined to minimise the impact of the inpainting procedure on the $E$-mode map statistics.

We find that statistical assessments of the $E$-mode data alone do not provide any evidence for or against the fluke hypothesis. However, tests based on
cross-statistical measures determined from temperature and $E$ modes
can allow this hypothesis to be rejected with a moderate level of
probability.

}

\begin{document}
\maketitle
\flushbottom


\newpage

\section{Introduction}
\label{sec:introduction}

\LB\ (the Lite spacecraft for the study of $B$-mode polarization and Inflation from cosmic background Radiation Detection) is a strategic large-class mission selected by the Japan Aerospace Exploration Agency (JAXA) to be launched in the 2030s. This work is part of a series of papers that present the science achievable by the \lb\ space mission, expanding on the overview
published in ref.~\cite{Hazumi2021PTEP}. 

Observations of the cosmic microwave background (CMB) are remarkably consistent with the $\Lambda$CDM cosmological model, which is defined by only six parameters \cite{P18_VI_Params}. However, some features in the data are in mild tension with the best-fit model, and could hint at new physics to be explored. These features, often referred to as ``anomalies", have been found in both the \wmap\ and \Planck\ temperature maps, albeit at modest levels of statistical significance  (i.e., $2$--$3\,\sigma $), as described in 
refs.~\cite{P13_XXIII_InS,P15_XVI_InS,P18_VII_InS} and references therein. The most important anomalous statistical signatures are generally identified as: a low-$\ell $ power deficit (low variance); a lack of correlation on large angular scales in the angular 2-point correlation function; alignment between the quadrupole and octopole moments; a hemispherical asymmetry (or dipolar modulation) of power; a parity asymmetry defined by the spherical harmonic modes with even and odd $\ell$ values respectively; and an anomalous ``Cold Spot" on angular scales of approximately $10^\circ$. 

The simplest explanation for the anomalies, given their claimed levels of significance, is that they are simply statistical excursions, caused by mildly unusual patterns in the temperature anisotropies arising in the standard cosmological model; we refer to this as the \lq\lq fluke hypothesis." 
However, since the temperature fluctuations are almost cosmic-variance limited, new observations must be sought in order to assess whether this is the case or 
there is a more interesting explanation.
Fortunately, maps of the polarised CMB anisotropies independently probe the fluctuations that
source the temperature field, thus providing exactly such information, and it is expected that much progress will be possible using \LB\ data.
Additional strong motivation for undertaking such studies with the
\lb\ data is that the \lq\lq a posteriori statistics" issue with the
anomalies is avoided. Since many tests could be applied to the data to
search for anomalous behaviour, we expect some outliers at the
approximately 3\,$\sigma$ level. Deciding how to interpret such levels of
significance will then exhibit a degree of subjectivity.  However, if
a fixed set of statistical tests that indicate potential anomalies in
temperature are applied to independent polarisation data, then
the problem of a posteriori or data-driven choices being made in the
analyses is avoided.
Of course, given the modest significance of the temperature features, high significance detections of anomalies in polarisation will still prove difficult.

Indeed, no definitive evidence was found in the \Planck\ 2018 polarisation data \cite{P18_VII_InS} for any anomalous features corresponding to those observed in the temperature
data. However, the various tests of isotropy were clearly limited by the presence of instrumental noise as well as residual systematics, even if these did not dominate the signal as was the case for the 2015 data set \cite{P15_XVI_InS}. This was clearly evident from the variation in results found with the four component separated maps studied, presumably related to their different responses to noise and systematic residuals, and an incomplete understanding of the noise properties of the data.  
In the case of \LB, the $E$-mode signal in particular is expected to be measured at close to the cosmic variance level, and the final sensitivity defined almost entirely by the sky fraction available for analysis.

However, no theoretical model of primordial perturbations has yet been constructed that can explain all of the temperature anomalies. Moreover, making inferences about the amplitude of anomalous features that might be observed in the \LB\ polarisation data based on what is seen in temperature is non-trivial. Specifically, predictions must be based on models constructed in 3D position space then propagated to spherical harmonic space. This mapping is different for temperature and polarisation.
An economical approach then, followed in this paper, is to focus an initial assessment on testing the consequences of the fluke hypothesis. 
This can be achieved by comparing the properties of a specific statistic in polarisation built from constrained simulations (where the part of the $E$-mode anisotropy correlated with temperature is constrained by current observations of the latter) with that constructed from unconstrained realisations. 
Similar studies have previously been presented in refs.~\cite{Copi:2013zja,Contreras2017,ODwyer2017,Shi2023}.

The intent of this paper is to optimise our ability to address the fluke hypothesis. Therefore, we work with simulations of the sky uncompromised by instrumental noise or residual foreground emission, and study their statistical properties over the largest sky fraction possible. Indeed, an important consideration in the search for anomalies is to ensure that the sky fraction under investigation provides a fair and unbiased sample of the CMB signal, thus minimal masking is preferred. Since many of the anomalous features seen in the CMB temperature data are on large angular scales, we generally study low-resolution simulations, such that our conclusions are not specific to \LB\ in detail. 
We also adopt a direct template of the CMB temperature anisotropy. 
The \planck\ 2018 component-separation study \cite{P18_IV_CS} provided four estimates of the CMB temperature anisotropy -- \commander, \nilc, \sevem, and \smica\ -- 
and here we use the latter. 
Preliminary tests indicated that the explicit choice of template did not affect our results significantly.
Interestingly, the analysis is then actually limited by the properties of that data, in particular, the impact of residual foreground emission predominantly towards the Galactic plane. Therefore, a so-called \lq\lq inpainting" method is used to replace the high-foreground regions with a Gaussian constrained realisation. Given that such data are then propagated into the constrained $E$-mode simulations, we test which analyses are sensitive to this processing and apply suitable masks where necessary.
For each applied test, we also verify that the temperature-correlated
$E$ modes generated from the \smica\ map exhibit statistical
excursions. This need not be the case, since they are produced by a
scale-dependent, non-local transformation of the temperature
data. Obviously, in the absence of such behaviour, a given test
will have no statistical power to evaluate the fluke hypothesis.

The paper is organized as follows.
Section~\ref{sec:poln_analysis} provides a brief introduction
to the study of polarised CMB data. Section~\ref{sec:simulations} describes the
characteristics of the simulations that constitute our reference set
of Gaussian sky maps, while section~\ref{sec:data} provides a prescription for generating inpainted realisations of the temperature data based on \planck\ observations 
that are then used to constrain the $E$-mode simulations.
Sections~\ref{sec:moments} through \ref{sec:large_scale_peaks} address specific statistical tests of the simulated polarisation data, motivated by previously observed anomalous statistical features of the microwave sky. Finally, section~\ref{sec:conclusions} details the main conclusions of the paper.

\section{Polarisation analysis preamble}
\label{sec:poln_analysis}

The measured quantities describing the linear polarisation state of the CMB are the standard $Q$ and $U$ Stokes parameters \cite{Zaldarriaga1997}. However, since these form the components of a rank-2 polarisation tensor in a specific coordinate basis associated with the map, they are not rotationally invariant. Alternatively, a global analysis of CMB data can be undertaken using the rotationally invariant $E$ and $B$ modes, defined as follows.

The quantities $Q\pm iU$, defined relative to the direction vectors $\hat{n}$
can be expanded as
\begin{equation}
(Q \pm iU) (\hat{n}) = \sum_{\ell = 2}^{\infty} \sum_{m=-\ell}^{\ell}a_{\ell m}^{(\pm 2)} \,_{\pm 2}Y_{\ell m}(\hat{n}),
\end{equation}
where $\,_{\pm 2}Y_{\ell m}(\hat{n})$ denotes the spin-weighted
spherical harmonics and $a_{\ell m}^{(\pm 2)}$ are the corresponding
harmonic coefficients. If we define
\begin{equation}
\begin{array}{ccc}
a^{E}_{\ell m} & = & -\dfrac{1}{2}\left(a_{\ell m}^{(2)} + a_{\ell m}^{(-2)}\right), \\[\medskipamount]
a^{B}_{\ell m} & = & \phantom{-}\dfrac{i}{2}\left(a_{\ell m}^{(2)} - a_{\ell m}^{(-2)}\right),
\end{array}
\end{equation}
then the invariant quantities are given by
\begin{equation}
\begin{array}{ccc}
E(\hat{n})& = & \displaystyle \sum_{\ell = 2}^{\infty} \sum_{m=-\ell}^{\ell} a^{E}_{\ell m} Y_{\ell m}(\hat{n}), \\
B(\hat{n}) & = & \displaystyle \sum_{\ell = 2}^{\infty} \sum_{m=-\ell}^{\ell} a^{B}_{\ell m} Y_{\ell m}(\hat{n}).
\end{array}
\end{equation}
In this paper, we focus on the $E$-mode signal given its partial correlation with the temperature anisotropies that have demonstrated interesting anomalous statistical features.

In addition, the quantities $Q_r$ and $U_r$, 
corresponding to a local rotation of the Stokes parameters,
are employed in section~\ref{sec:large_scale_peaks}.
In this case,
a local frame is defined with respect to a reference point $\hat{n}_{\rm{ref}}$.
The new quantities are then calculated as \cite{P18_VII_InS}
\begin{equation}
    \begin{pmatrix}
    Q_r(\hat{n},\hat{n}_{\rm{ref}}) \\
    U_r(\hat{n},\hat{n}_{\rm{ref}}) 
    \end{pmatrix}
    =
    \begin{pmatrix}
    -\cos(2\phi) & -\sin(2\phi) \\
    \phantom{-}\sin(2\phi)  & -\cos(2\phi) 
    \end{pmatrix}
    \begin{pmatrix}
    Q(\hat{n}) \\
    U(\hat{n}) 
    \end{pmatrix} \, ,
    \label{eq:Qr_Ur}
\end{equation}
where $\phi$ is the angle between the axis aligned along a meridian in the local coordinate system centred on the reference point and the great circle connecting this point to a position $\hat{n}$.

\section{Simulations}
\label{sec:simulations}

The results presented in this paper are derived using Monte Carlo (MC) simulations of the polarised CMB sky. 
We make extensive use of both random isotropic (or unconstrained) realisations of a fiducial cosmological model, and constrained simulations where the temperature map is based on the 
\Planck\ \smica\ PR3 data, available  on the Planck Legacy Archive
(PLA\footnote{\url{http://pla.esac.esa.int}}).  

The fiducial cosmological model corresponds to the \Planck\ 2018 baseline marginalised results \citep{P18_VI_Params}, with the six $\Lambda$CDM parameters 
$H_{0}$ (Hubble constant), 
$\Omega_{\rm b}h^{2}$ (baryon density), 
$\Omega_{\rm c}h^{2}$ (dark matter density), 
$\tau$ (reionisation optical depth), 
$n_{\rm s}$ (scalar spectral index), 
and $A_{\rm s}$ (scalar amplitude)
corresponding to values of 
$67.36\ {\rm km}\,{\rm s}^{-1}\,{\rm Mpc}^{-1}$, 
$0.0224$, $0.1202$, $0.0544$, $0.9649$, and $2.099\times 10^{-9}$ respectively,
including cosmological lensing but no tensor perturbations, i.e., $r=0$.

The \smica\ sky map corresponds to one of the four estimates of the CMB sky derived from the component-separation algorithms applied to the \planck\ 2018 data \cite{P18_IV_CS}. These algorithms combine the \planck\ frequency maps to yield data with minimal Galactic foreground residuals. However, some residuals do remain, predominantly in the Galactic plane, and we make use of an
inpainting technique, described in section~\ref{sec:data}, to 
replace all high-foreground regions with a Gaussian constrained realisation. Since this method is not deterministic, $1200$ inpainted \smica\ realisations are used as constraints for the $E$-mode simulations. For consistency, the unconstrained temperature simulations are also inpainted, although only a single inpainting realisation is used for each simulated sky. 

For further consistency between the temperature simulations and data, \smica\ noise realisations are added to the former before inpainting. $300$ so-called FFP10 full focal-plane noise simulations that capture the complex instrumental noise properties and residual systematic effects of the \planck\ detectors are available from the PLA. 
These are extensions of the simulations described
in ref.~\cite{P15_XII_FFP}.
The noise simulations are permuted with $1200$ signal realisations to generate the final simulated data set.

The inpainting of the \smica\ data, and the need to treat the unconstrained simulations consistently, necessarily has implications for the 
corresponding $E$-mode signal realisations, which comprise a part that is correlated with the temperature anisotropy, and an uncorrelated part. This can be written in terms of spherical harmonic coefficients as \cite{constrained_e_simulations}
\begin{equation}
    a^{E}_{\ell m} = \dfrac{C^{TE}_{\ell}}{C_{\ell}^{TT}}a^{T}_{\ell m} + \sqrt{\dfrac{C_{\ell}^{TT}C_{\ell}^{EE}-(C_{\ell}^{TE})^{2}}{C_{\ell}^{TT}}}g_{\ell m}\, ,
\label{eq:alm_T_and_E}
\end{equation}
where the $g_{\ell m}$ are values sampled from a complex standard normal distribution and, $C_{\ell}^{TT}$, $C_{\ell}^{EE}$, and $C_{\ell}^{TE}$ are defined by the fiducial CMB power spectra. 
An $E$-mode realisation is then constructed by setting the 
temperature anisotropy coefficients to the values obtained from the inpainted temperature map.

Note that eq.~(\ref{eq:alm_T_and_E}) indicates why, in part,
we use inpainted rather than masked temperature sky maps. Specifically, part of the $E$-mode signal
is based on a non-local and scale-dependent transformation of the temperature signal, and masking may introduce artefacts.

In particular, we generate and analyse sky maps in {\tt HEALPix}\footnote{\url{http://healpix.sourceforge.net}} format \cite{gorski2005}, with a pixel size defined by the \nside\ parameter.\footnote{In {\tt HEALPix} the sphere is divided into $12\,N_{\rm side}^2$ equal area pixels. At $N_{\rm side}=(512, 64)$, the mean pixel size is  $(6.9^\prime, 55^\prime)$.}
Although we anticipate that the typical CMB product from \LB\ after component separation will likely correspond to a full width at half maximum (FWHM) angular resolution of $30^\prime$ and \nside\ = $512$, in this work we analyse low-resolution simulations generated at 
\nside\ = $64$ with a corresponding FWHM angular resolution of $160^\prime$.  
Since the \smica\ sky map and FFP10 noise simulations are provided at high resolution, they are first degraded to lower resolution as follows. The full-sky maps are decomposed into spherical harmonics at
the input \healpix\ resolution. These coefficients are then effectively convolved
to the new resolution by rescaling them using the appropriate ratio of output and input beam and pixel window functions, then the modified coefficients are used to synthesise a map
directly at the output \healpix\ resolution. 
Section~\ref{sec:moments} also describes a study of lower resolution simulations and data at \nside\ $\in \{8, 16, 32\}$ with a corresponding FWHM angular resolution of $\{1280^\prime, 640^\prime, 320^\prime\}$, respectively, generated following this approach.
These \nside\ and FWHM choices correspond to those adopted for low-resolution studies in ref.~\cite{P18_VII_InS}.
Exceptionally, the analyses presented in section~\ref{sec:large_scale_peaks} need to be undertaken at a higher resolution, and independently generated simulations are utilised, see section~\ref{subsec:LSP_simulations}.

\section{Inpainting}
\label{sec:data}

\begin{figure}
    \begin{subfigure}[t]{.5\linewidth}
        \centering
        \includegraphics[width=\linewidth]{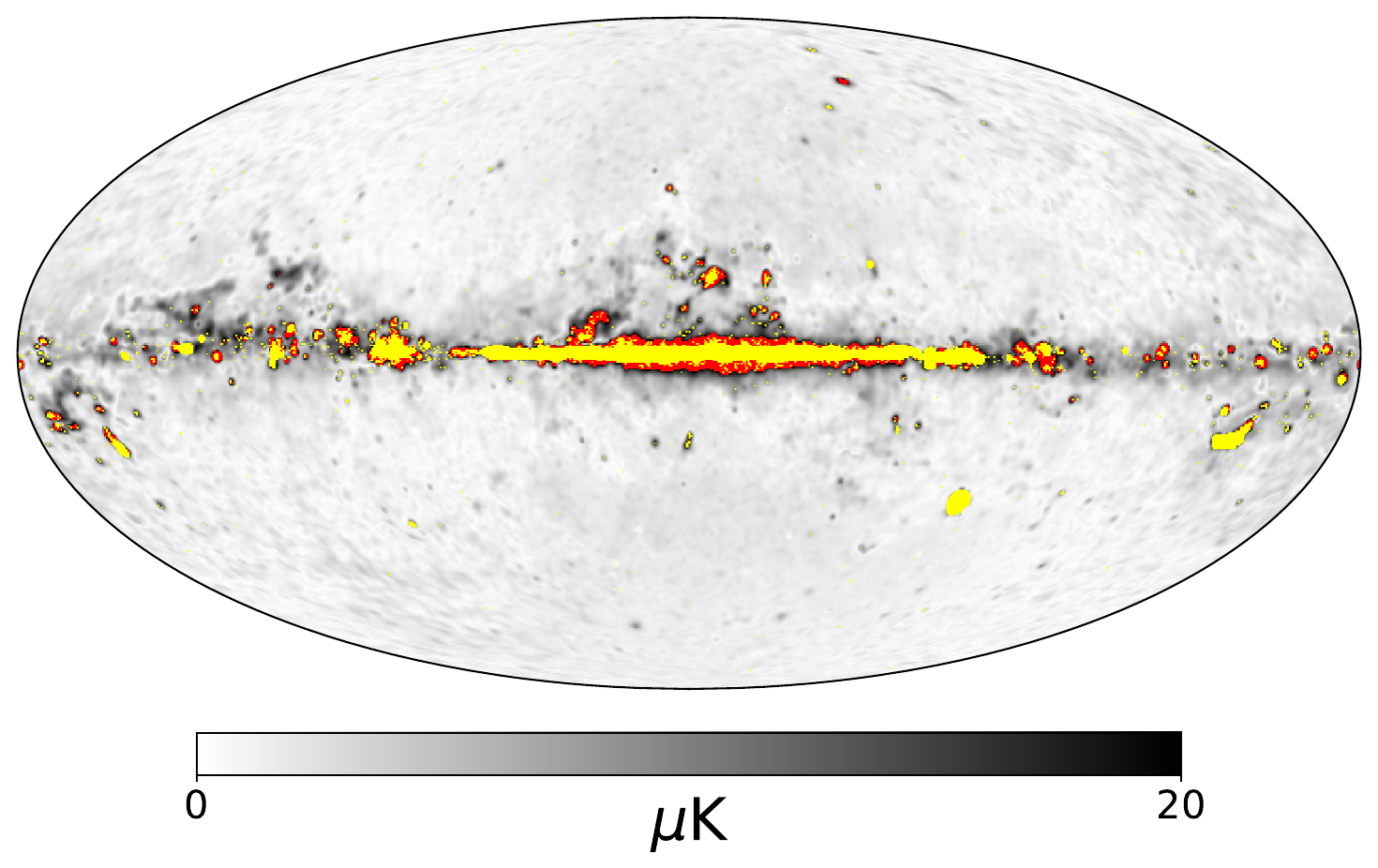}
        \label{}
    \end{subfigure}
     \begin{subfigure}[t]{.5\linewidth}
        \centering
        \includegraphics[width=0.9\linewidth]{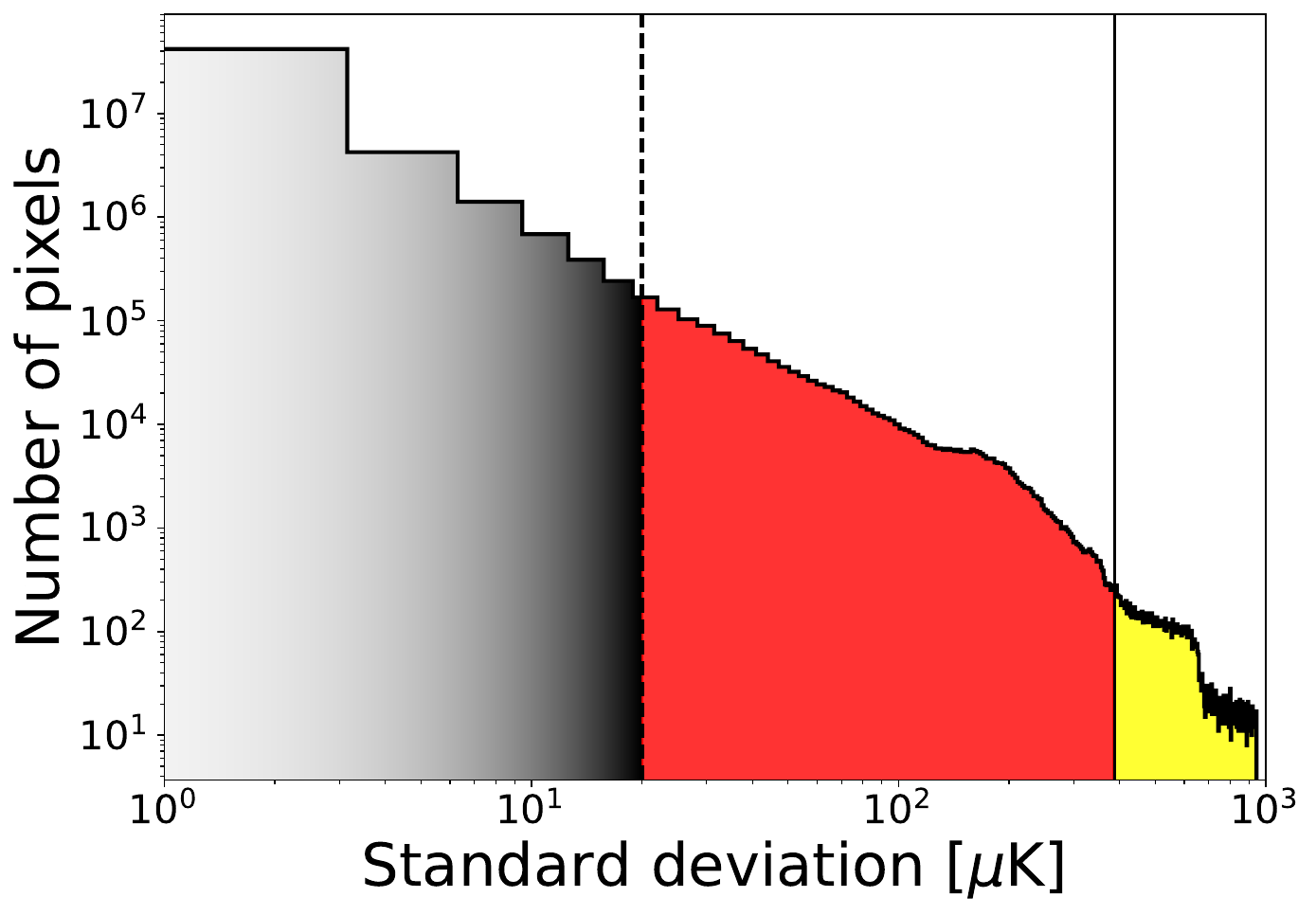}
        \label{}
    \end{subfigure} 
    \caption{\textit{Left:} map of the standard deviation per pixel determined from the four component-separation methods after monopole and dipole subtraction and smoothing to $80^\prime$ resolution. The yellow area corresponds to the original inpainting mask, while the red region shows the extension defined by applying a threshold of 20 $\mu$K. \textit{Right:} histogram of the standard deviation map.
    The solid black vertical line indicates the 380 $\mu$K threshold defining the original \planck\ 2018 inpainting mask, while the black dashed vertical line corresponds to the 20 $\mu$K threshold defining the inpainting mask used here. The same colours from the left panel are used to identify the original \planck\ 2018 inpainting mask and the extended version. The region below the 20 $\mu\rm{K}$ theshold corresponds to the final unmasked sky fraction of $96.9\%$.}
    \label{fig:Std_Map}
\end{figure}

\begin{figure}
    \centering
    \includegraphics[width=15cm]{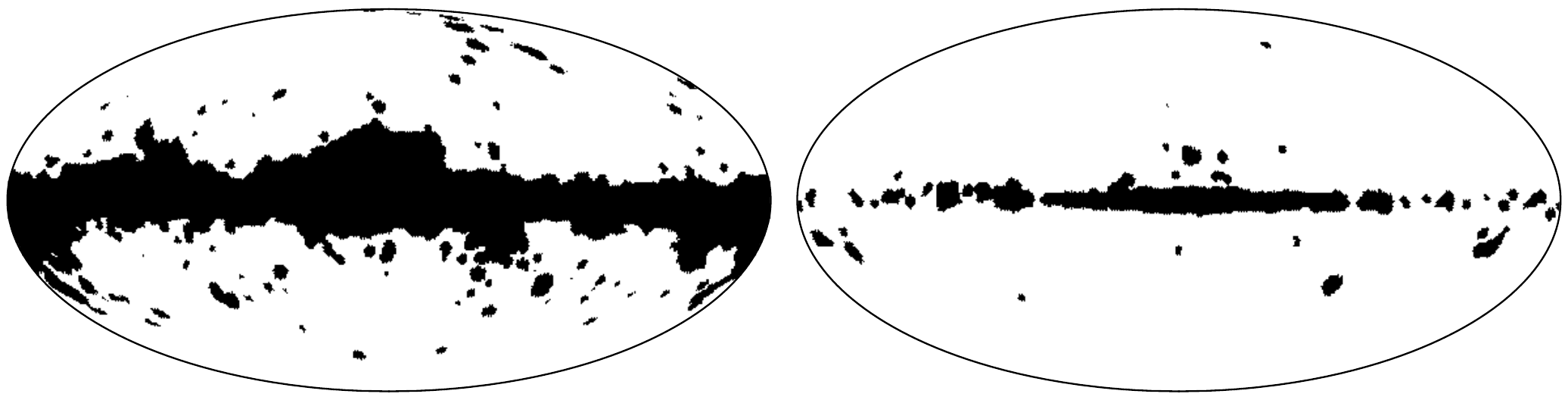} 
    \caption{Masks used to inpaint the \Planck\ PR3 \smica\ data at \nside\ = 64. \textit{Left:} common confidence mask for temperature. \textit{Right:} inpainting mask for temperature. Both are downgraded to \nside\ = 64, leaving 71.3\% and 93.9\% of the sky available for analysis, respectively.}
    \label{fig:masks}
\end{figure}

As noted previously, we use the \planck\ \smica\ data as a template for the CMB temperature field, but since highly foreground-contaminated regions that are usually masked for analysis are present, and we require a full sky temperature map to constrain the $E$-mode polarisation simulations,  
an inpainting approach, \texttt{CMB-PAInT}~\cite{Gimeno-Amo2024}, is applied to replace the masked areas by a Gaussian constrained realisation (GCR). 
In particular, masked pixels are filled by sampling from a conditional probability distribution defined by a fiducial sky model and the values of the unmasked pixels. 

The \planck\ 2018 component-separation study \cite{P18_IV_CS} defined a common mask for the analysis of the four estimates of the CMB temperature anisotropy (\commander, \nilc, \sevem, and \smica) at \nside\ = $2048$ and $5^\prime$ resolution. Lower resolution versions were generated in a conservative manner by appropriately smoothing, degrading, and thresholding the mask. We consider the \nside\ = 64 common mask as one option to mask the low resolution \smica\ data before inpainting. However, since the full-sky information after inpainting will depend on the mask, it is potentially advantageous to consider one that retains a larger sky fraction. An inpainting mask was also defined in ref.~\cite{P18_IV_CS}, but not rigorously derived and the resulting inpainted data intended only for visualisation purposes. 
Since the GCR approach is very sensitive to any mismatch between the fiducial model and the values of the unmasked pixels, and given the possibility that residual foregrounds are still present in the data when masked on such a small sky fraction, a dedicated low-resolution inpainting mask is derived here based on the four \planck\ 2018 component-separated maps.

Initially, for each of the component-separated temperature maps at high resolution, the best-fit dipole and monopole computed when applying the original inpainting mask is removed. Then, 
following the methodology used to define the common mask at high resolution,
the four maps are smoothed to $80^\prime$ resolution and the standard deviation is computed for each pixel. Figure~\ref{fig:Std_Map} shows the standard deviation map. The original \planck\ 2018 inpainting mask is then extended by rejecting all pixels with a standard deviation value larger than 20 $\mu$K.  
This is then smoothed to an effective resolution of $160^\prime$, and an \nside\ = 64 mask generated by applying a threshold of 0.9, such that low-resolution pixels below this value are rejected, and the remainder reset to unity, resulting in a binary mask with 93.9\% usable sky coverage. Hereafter, this is referred to as the inpainting mask.

Lower resolution common masks at $N_\mathrm{side}=\{8, 16, 32, 64\}$ are derived similarly, after smoothing the original common mask at $N_\mathrm{side}=2048$ to an effective resolution of $\mathrm{FWHM}=\{1280', 640', 320', 160'\}$ respectively,  then applying a 0.9 threshold to allow the creation of a binary mask. The resulting usable sky coverage at each resolution is $f_\mathrm{sky}=\{55.2\%, 64.5\%, 68.8\%, 71.3\%\}$. \

\begin{figure}[t!]
    \centering
    \includegraphics[width=15cm]{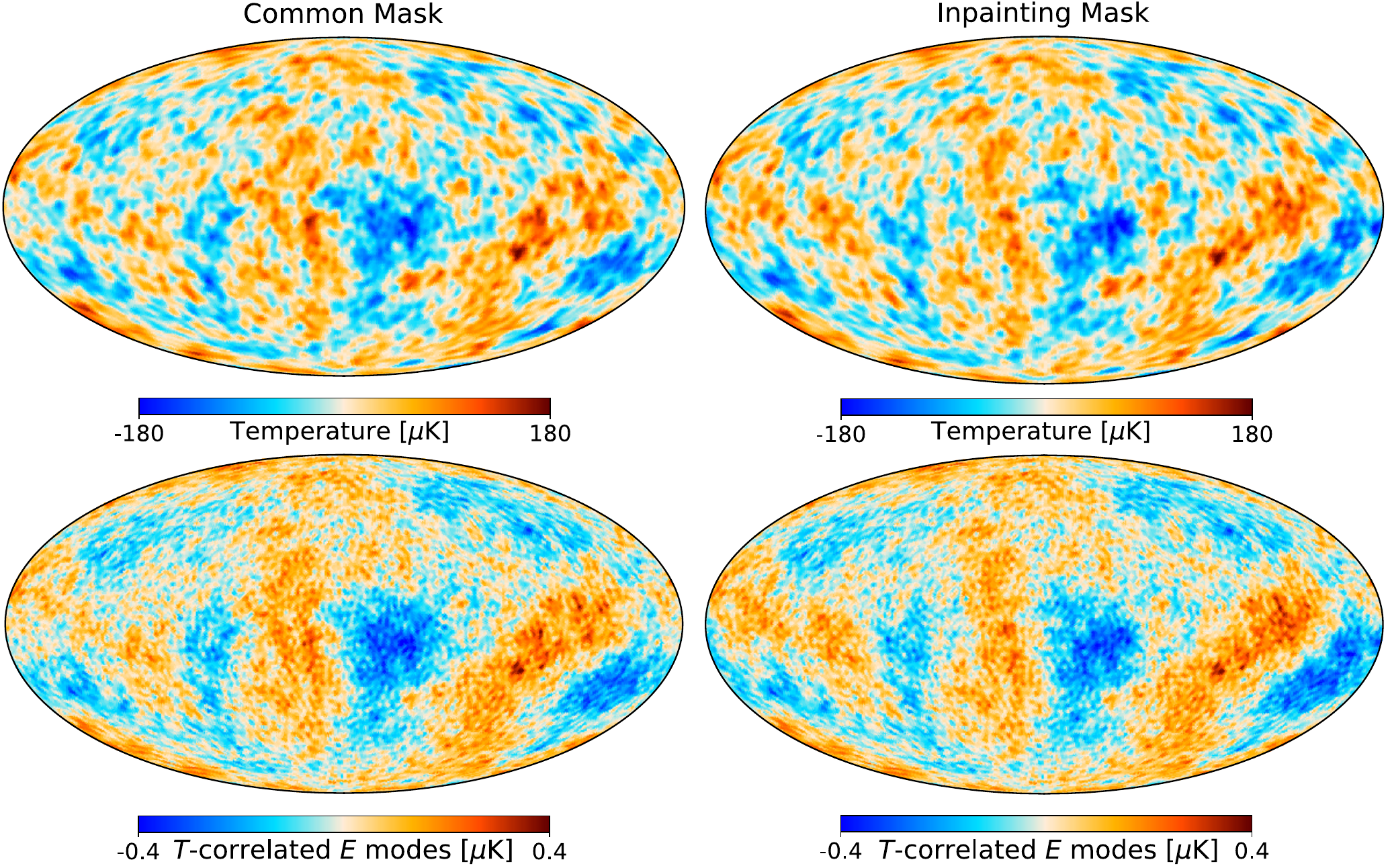} 
    \caption{\textit{Top:} example realisations of the \smica\ temperature data inpainted on the common and inpainting masks. \textit{Bottom:} the corresponding temperature-correlated part of the $E$-mode sky map.}
    \label{fig:Inpainted_maps}
\end{figure}

The \smica\ data are then inpainted at \nside\ = $64$ using two different masks, the \Planck\ common mask and the inpainting mask described above, shown in Figure~\ref{fig:masks}. 
Since the inpainting process is stochastic, 1200 realisations are generated and used to constrain $E$-mode simulations. Figure~\ref{fig:Inpainted_maps} presents one inpainted realisation of the data for each mask. Note that, to be fully consistent with this processing, the unconstrained simulations also utilise inpainted realisations of temperature, although only one per simulated sky. 
The data and simulations at lower resolutions are directly inpainted at these resolutions using the appropriate masks.

In the analyses that follow, we remain cognisant of the fact that inpainting is not a remedy for contaminated data, and that the resulting full-sky maps are dependent on method assumptions and the applied mask. 
Under the fluke hypothesis, the CMB temperature sky is considered to be consistent with the $\Lambda$CDM cosmological model. However, it is statistically unusual, so that when the inpainted regions are filled with values from an arbitrary realisation of the standard cosmological model, the process could potentially dilute the significance of anomalies computed from full sky data.
We therefore test whether the full sky inpainted \smica\ temperature maps, and the temperature-correlated part of the $E$-mode data derived therefrom, are anomalous on the full sky, and otherwise apply appropriate masking. 
In fact, we find that only the alignment statistic presented in section~\ref{sec:lowl_alignment} is sufficiently insensitive to the inpainting process to allow full sky analysis. This is consistent with previous studies using this statistic that have applied  inpainting methods of various types to different sky cuts \cite{P13_XXIII_InS,Starck:2013,Copi:2015}.
It should also be noted that some anomalies become increasingly significant with decreasing sky coverage, and in such cases the use of masks is again preferred. 

\begin{figure}[t!]
    \centering
    \includegraphics[width=14cm]{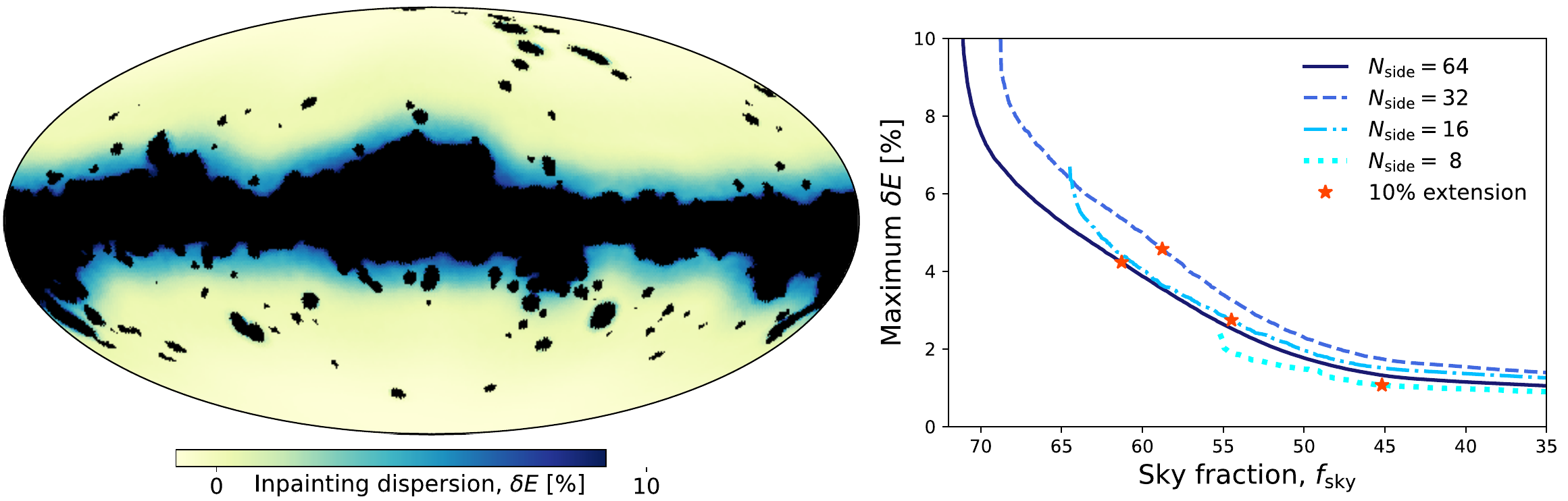}
    \caption{Extension of \Planck's temperature common mask to $E$-mode polarisation. \textit{Left:} dispersion that the inpainting of $N_\mathrm{side}=64$ temperature maps introduces on $E$-mode maps through the $TE$ correlation. \textit{Right:} maximum dispersion as a function of the sky coverage allowed by \Planck's temperature common mask at different resolutions. In this work, we define $E$-mode masks by reducing by $10\%$ the sky coverage of the initial temperature mask. Pixels with $\delta E$ above that threshold will be masked.}
    \label{fig:mask extension to E modes}
\end{figure}

\begin{figure}
    \centering
    \includegraphics[width=15cm]{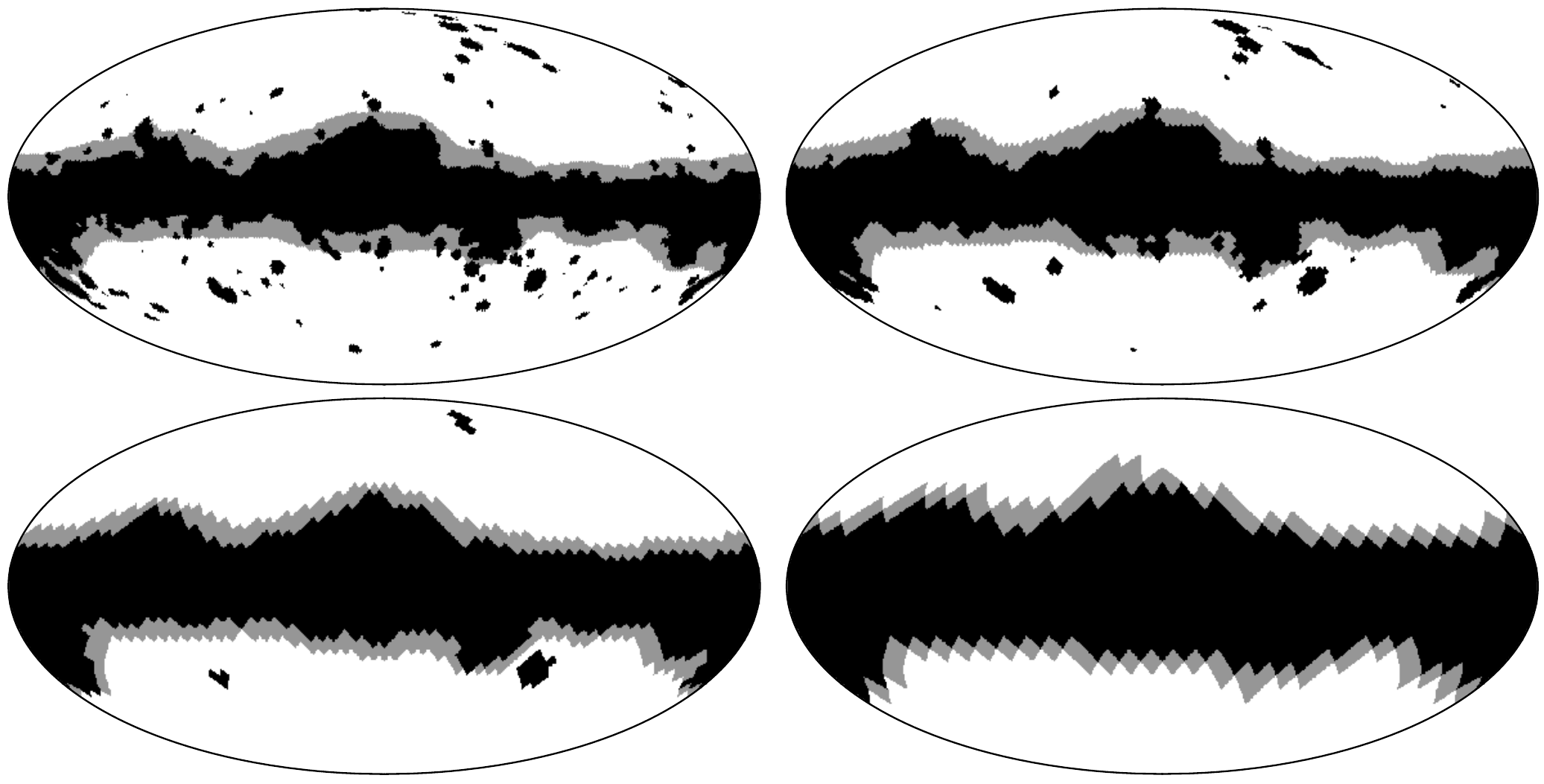} 
    \caption{\Planck\ common temperature mask (black) and its extension to $E$-mode polarisation (grey). From left to right, top to bottom, we show $N_\mathrm{side}=64, 32, 16, 8$ resolutions. The corresponding sky fractions are, respectively, $f_\mathrm{sky}= 71.3\%, 68.8\%, 64.5\%, 55.2\%$ for temperature and $f_\mathrm{sky}= 61.3\%, 58.8\%, 54.5\%, 45.2\%$ for polarisation.}
    \label{fig:low resolution masks}
\end{figure}

\begin{figure}
    \centering
    \includegraphics[width=8cm]{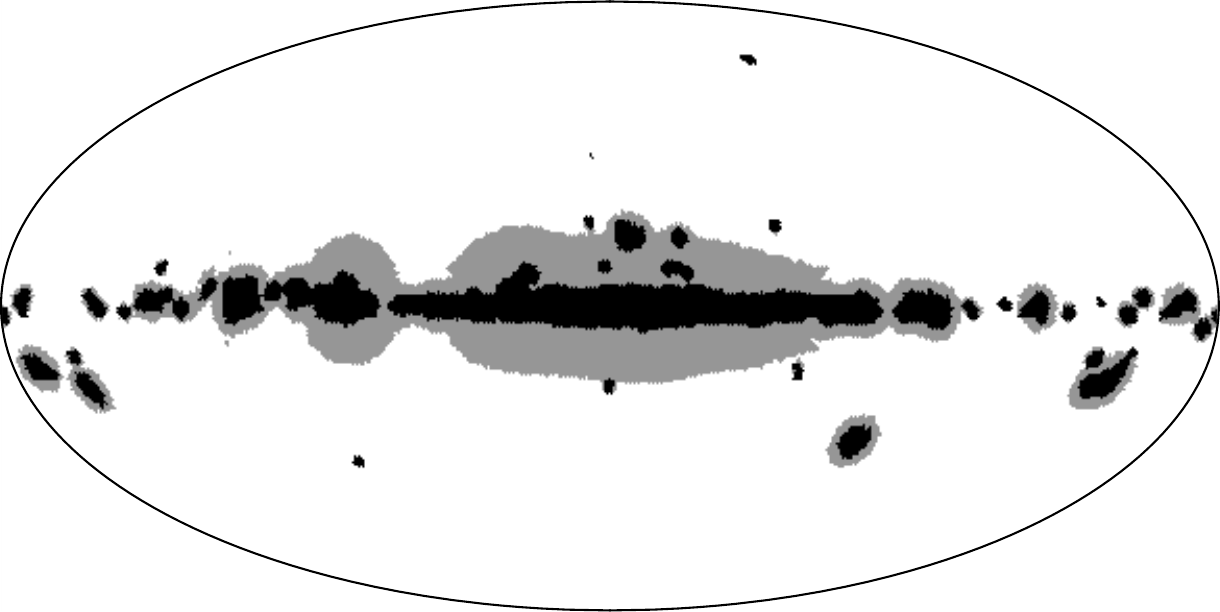} 
    \caption{Inpainting temperature mask (black) and its extension to $E$-mode polarisation (grey) at $N_\mathrm{side}=64$ resolution. The corresponding sky fractions are $f_\mathrm{sky}= 93.9\%$ for temperature and $f_\mathrm{sky}= 83.9\%$ for polarisation.}
    \label{fig:E-mode extended inpainting mask}
\end{figure}

We therefore derive masks for the analysis of the $E$-mode sky maps by consideration of the properties of the temperature-correlated part of the polarised signal in our 1200 unconstrained simulations. Specifically, we determine the difference between the $E$-mode polarisation map constrained to the original and inpainted  versions of the temperature sky for each realisation, $\Delta_i (\vec{n}) = E_{\mathrm{org},i}(\vec{n}) - E_{\mathrm{inp},i}(\vec{n})$, then define the 
dispersion introduced by the inpainting as 
\begin{equation}
    \delta E (\vec{n}) = \sqrt{ \frac{\sum_i \left(\Delta_i (\vec{n}) - \bar{\Delta} (\vec{n})\right)^2}{\sum_i \left(E_{\mathrm{org},i} (\vec{n}) - \bar{E}_\mathrm{org} (\vec{n})\right)^2} }. 
\end{equation}
Figure~\ref{fig:mask extension to E modes} shows the $\delta E$ dispersion outside the common mask at $N_\mathrm{side}=64$ resolution. The common mask is then extended by masking an additional $10\%$ fraction of the pixels with the largest $\delta E$. We adopt a $10\%$ extension of the temperature mask as a compromise between preserving the anomalous nature of the \texttt{SMICA} temperature data and retaining sufficient sky coverage. 
The $E$-mode common masks at $N_\mathrm{side}=8,16,32,64$ are
shown in figure~\ref{fig:low resolution masks}.  The same procedure is applied to produce the $E$-mode version of the inpainting mask at $N_\mathrm{side}=64$ resolution, shown in figure~\ref{fig:E-mode extended inpainting mask}.

\section{Tests of non-Gaussianity: one-dimensional moments}
\label{sec:moments}

The standard cosmological model predicts that an early phase of accelerated expansion gave rise to density perturbations distributed as a statistically homogeneous and isotropic Gaussian random field. The same statistical properties were imprinted on the primordial CMB according to linear theory. In this section, we look for signs of non-Gaussianity and deviations from $\Lambda$CDM in the anisotropies of the $E$-mode polarisation using estimators of its second-, third-, and fourth-order moments. 

We use the following definitions for the variance, skewness, and kurtosis of any $w,x,y,z$ variables:
\begin{align}
\sigma_{xy}^2= & \frac{1}{N} \sum\limits_i (x_i - \bar{x}) (y_i - \bar{y}); \label{eq: cov xy}\\
S_{xyz}= & \frac{1}{N} (\sigma_{xx}\sigma_{yy}\sigma_{zz})^{-1} \sum\limits_i (x_i - \bar{x}) (y_i - \bar{y}) (z_i - \bar{z}); \label{eq: skew xyz}\\
k_{wxyz}= & \frac{1}{N} (\sigma_{ww}\sigma_{xx}\sigma_{yy}\sigma_{zz})^{-1}\sum\limits_i (w_i - \bar{w}) (x_i - \bar{x}) (y_i - \bar{y}) (z_i - \bar{z}). \label{eq: kurt wxyz non-central}
\end{align}
Here $\bar{x}=N^{-1} \sum_i x_i$.  A normal distribution has null skewness according to eq.~(\ref{eq: skew xyz}). Eq.~(\ref{eq: kurt wxyz non-central}) describes the standardised fourth-order moment, for which a normal distribution would give $k=3$. However, for a Gaussian field as strongly correlated as the CMB, a very large number of samples is needed to overcome the correlation between second- and fourth-order moments and recover this value. Such an ergodicity condition is not fully satisfied with the reduced number of pixels contained in any of the low-resolution maps studied in this section. Therefore, we de-bias our kurtosis statistic by subtracting the average value determined from the unconstrained $\Lambda$CDM realisations,
\begin{equation}
K_{wxyz} = k_{wxyz} -  \langle  k_{wxyz} \rangle_\mathrm{unconstrained}. \label{eq: kurt wxyz central}
\end{equation}

Analyses of \WMAP\ \cite{Monteserin:2007fv, Cruz:2010ud} and \Planck\ \cite{P13_XXIII_InS, P15_XVI_InS, P18_VII_InS} temperature data report an anomalously low variance as compared to $\Lambda$CDM at larger angular scales. Thus, we perform a multi-resolution analysis of the \texttt{SMICA}-constrained simulations to progressively isolate the effect of the lack of power observed at the lowest multipoles. Moreover, previous work has shown that this anomalous behaviour is very mask dependent, and increases in significance with decreasing sky fraction \cite{Cruz:2010ud, Gruppuso:2013xba,P15_XVI_InS}.\footnote{As a reference, we find that the significance of the low-variance anomaly decreases, with a $p\mathrm{-value}$ of 1.75\% increasing to 8.42\% and 11.42\% when the \smica\ temperature map at $N_\mathrm{side}=64$ is analysed with, respectively, the \planck~common mask, our inpainting mask, and the original \smica~inpainting mask from ref.~\cite{P18_IV_CS}.} Therefore we mask the $E$-mode maps before calculating their real-space $n$-order moments to ensure that the anomalous nature of the \smica\ temperature anisotropies is retained in the correlated part of the $E$-mode anisotropies. For that purpose, we extend the \planck\ common mask defined in temperature following the procedure described in section~\ref{sec:data}. When analysed with a $10\%$ extension of the common mask, the temperature-correlated part of the $E$-mode maps constrained to the inpainted \texttt{SMICA} temperature data show an anomalously low variance at a $\geq 2.4\,\sigma$ level for all resolutions.

\begin{figure}
    \centering
    \includegraphics[width=\textwidth]{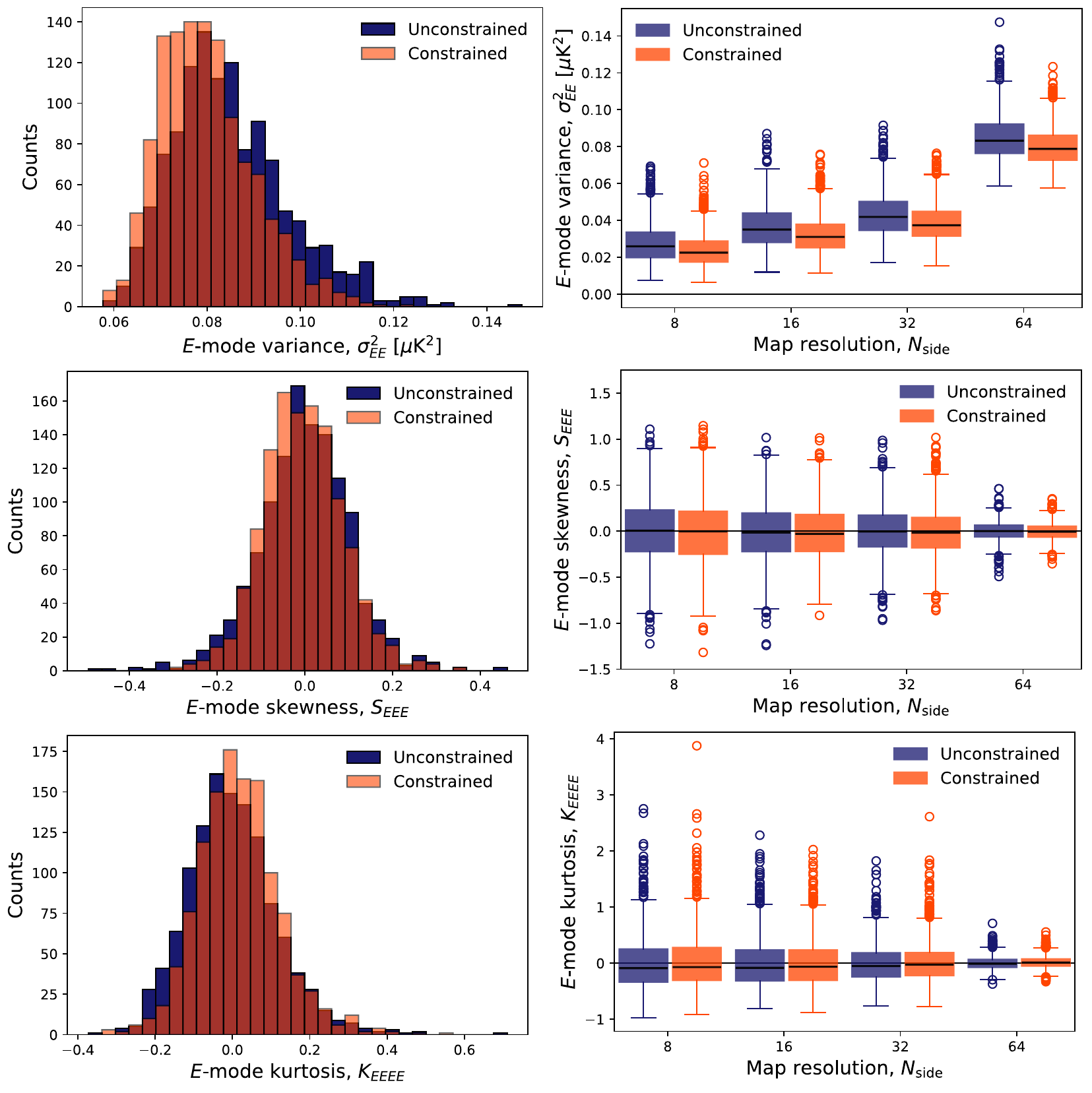}
    \caption{Distribution of the variance (top), skewness (centre), and kurtosis (bottom) of masked $E$-mode maps from constrained (orange) and unconstrained (blue) simulations, with the overlap shown in red.  \textit{Left:} distribution of $n$-order moments from simulations at $N_\mathrm{side}=64$ resolution. \textit{Right:} box-and-whisker plots comparing the distribution of moments at different resolutions. In these plots, boxes encompass the interquartile range $\mathrm{IQR} = Q_{0.75} - Q_{0.25}$, with a solid horizontal line to indicate the median ($Q_{0.50}$), and whiskers extending to $1.5\,\mathrm{IQR}$. Outliers beyond that range are represented by open circles.}
    \label{fig:Evar_Eskew_Ekurt}
\end{figure}

In figure~\ref{fig:Evar_Eskew_Ekurt}, we compare the distribution of the $n$-order moments from unconstrained and constrained simulations. The left panel shows the results obtained at $N_{\mathrm{side}}=64$ resolution, while the right panel extends the results to lower resolutions. As the top panel of figure~\ref{fig:Evar_Eskew_Ekurt} shows, the pull towards lower $\sigma^2_{EE}$ values that \smica's low variance in temperature induces through the $TE$ correlation is not enough to distinguish between constrained and unconstrained simulations: for all of the resolutions considered, there is no significant difference between the variances of constrained and unconstrained simulations. In this way, $E$-mode anisotropies alone do not have enough sensitivity to either confirm or dismiss the fluke hypothesis. Conversely, finding an anomalously low or high $E$-mode variance in \LB\ data would provide independent evidence against $\Lambda$CDM, since the anomalously low variance found in temperature does not significantly influence the statistics of $E$-mode anisotropies. The skewness (central panel) and kurtosis (bottom panel) distributions of constrained simulations do not show any anomalous behaviour when compared to that of unconstrained realisations.

\begin{figure}
    \centering
    \includegraphics[width=\textwidth]{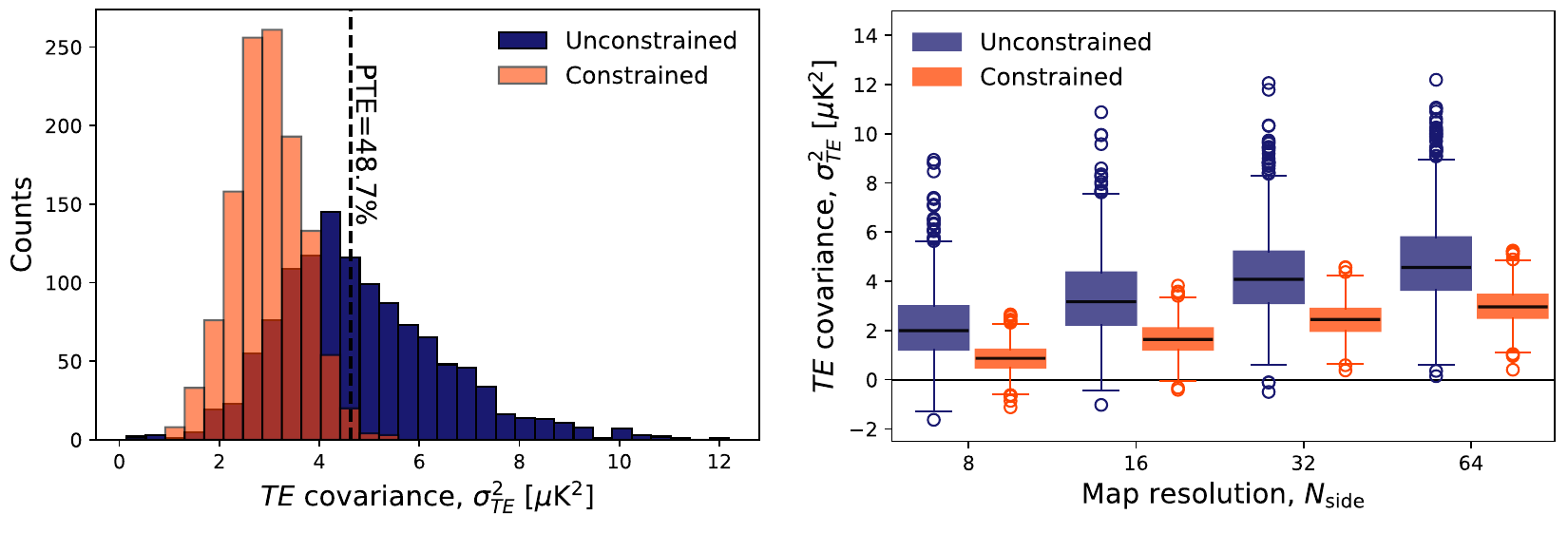}
    \caption{As figure~\ref{fig:Evar_Eskew_Ekurt}, but for the $TE$ covariance.}
    \label{fig:TEcovar}
\end{figure}

We can also perform a joint analysis of temperature and $E$-mode polarisation. In this case, we apply the $10\%$ extension of the common mask to both $T$ and $E$ maps. Figure~\ref{fig:TEcovar} shows the $\sigma^2_{TE}$ covariance obtained from constrained and unconstrained simulations at different resolutions. Now that temperature anisotropies are explicitly considered in the statistics, the anomalous properties of the \smica\ temperature data have an appreciable impact on the distribution of recovered values. Specifically, the low variance of the temperature anisotropy pulls the $TE$ covariance of constrained realisations towards lower values and reduces the width of the distribution in comparison to that of unconstrained simulations. As a result, at a resolution of $N_\mathrm{side}=64$, finding a $TE$ covariance above $4.62~\mu$K$^2$ would allow us to reject the fluke hypothesis at a $99\,\%$ confidence level with a $\mathrm{PTE}=48.7\,\%$. A similar $46$ to $51\%$ fraction of unconstrained simulations is found to have a larger $TE$ covariance than $99\,\%$ of constrained realisation for all resolutions. Therefore, there is a limited range of $\sigma^2_{TE}$ values in which future $E$-mode measurements will be compatible with \planck\ temperature data. Alternatively, we note that finding a $\sigma^2_{TE}\leq1.75~\mu$K$^2$ covariance in \LB\ data could also refute the fluke hypothesis, since it indicates that not only temperature but also $E$-mode anisotropies have an anomalously low variance compared to $\Lambda$CDM. However, we refrain from quoting a confidence level, since our simulation set is not big enough to provide an accurate representation of such low probability events.

\begin{figure}
    \centering
    \includegraphics[width=\textwidth]{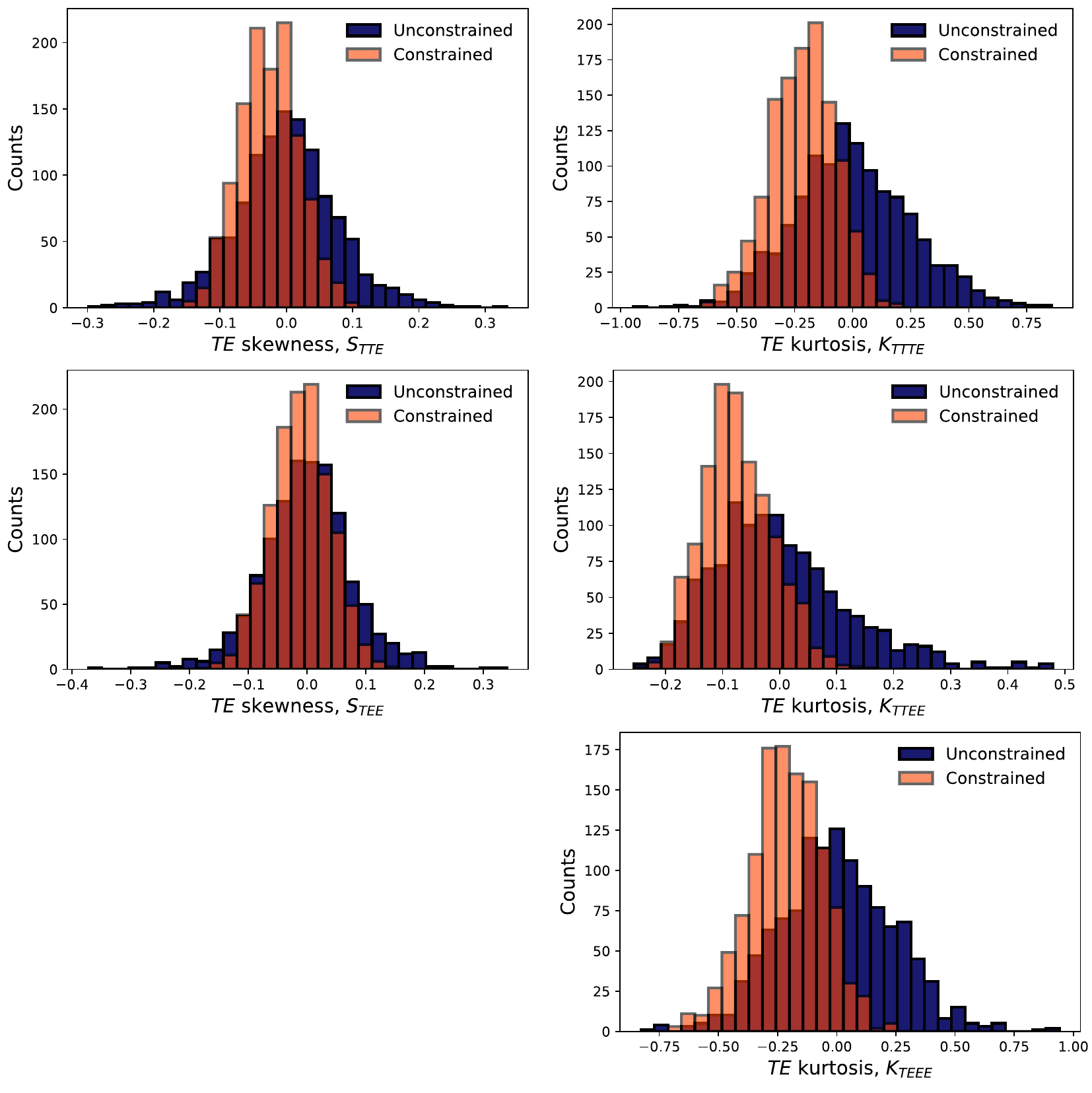}
    \caption{Skewness (left) and kurtosis (right) of different permutations of masked temperature and $E$-mode polarisation maps from constrained (orange) and unconstrained (blue) simulations at $N_\mathrm{side}=64$ resolution, with the overlap shown in red.}
    \label{fig:TE skew and kurt}
\end{figure}

Similarly, higher-order moments combining temperature and $E$-mode polarisation provide moderate sensitivity to ruling out the fluke hypothesis, as shown in figure~\ref{fig:TE skew and kurt} for the different permutations of skewness and kurtosis statistics. In particular, the sensitivity to the fluke hypothesis of the $K_{TTTE}$, $K_{TTEE}$, and $K_{TEEE}$ permutations of the fourth-order moment is, qualitatively, comparable to that of $\sigma^2_{TE}$, with constrained realisations favouring negative values. This suggests that, compared to the unconstrained pure $\Lambda$CDM simulations, \texttt{SMICA}-constrained realisations have fewer and less extreme outliers. In comparison, the $S_{TTE}$ and $S_{TEE}$ skewness permutations have less statistical power to rule out the fluke hypothesis. Nevertheless, the general behaviour is consistent with the kurtosis statistic: the presence of fewer and less extreme outliers in constrained realisations compared to $\Lambda$CDM simulations leads to a more symmetric distribution of values, with the skewness closer to zero.

These results afford evidence that a joint analysis of temperature and $E$-mode polarisation improves the power to reject the fluke hypothesis. Sensitivity to the fluke hypothesis could be maximised through a linear discriminant analysis combining different permutations of temperature and $E$-mode polarisation $n$th-order moments, a possibility that we leave for future work.

\section{Dipole modulation and directionality: variance asymmetry}
\label{sec:dipmod_varasym}

A hemispherical asymmetry, or dipolar modulation, of power was first discovered in the \textit{WMAP} first-year temperature data in refs.~\cite{Eriksen2004,Hansen2004,Park2004}. This anomaly, which indicates a deviation from the statistical isotropy of the standard cosmological model, was later confirmed through analyses of \textit{Planck} temperature data \cite{P13_XXIII_InS,P15_XVI_InS,P18_VII_InS}. In particular, several methods that are sensitive to either the amplitude or directionality of the modulation, or both, have been applied to \textit{Planck} temperature data \cite{P15_XVI_InS,P18_VII_InS}, 
yielding results consistent with a modulation of power of around $7\%$ between two hemispheres defined by a preferred direction $(l,b)=(209^\circ,-15^\circ)$ and extending over scales up to $\ell_{\rm max}\approx60$ with a significance approaching $3\,\sigma$. However, note that, in the absence of a theoretical model, there are debates on how to include corrections for a posteriori effects, as discussed, for example, in refs.~\cite{Bennett2011, P15_XVI_InS}. The study was then extended to \textit{Planck} polarisation data in ref.~\cite{P18_VII_InS}, and hints of an alignment between the preferred dipole modulation directions of the temperature and $E$-mode maps were found at  modest significance. Recently, similar results have been determined using the latest \textit{Planck} PR4 data set \cite{Gimeno-Amo_2023}. However, it was apparent that the various tests of isotropy applied to polarisation data were still limited by residual systematics.

Here, to test the fluke hypothesis, we will focus on a test sensitive to the dipolar modulation amplitude. In particular, we will look for a dipolar modulation of the pixel-to-pixel variance of the $E$-mode map, adopting the methodology presented in refs.~\cite{Akrami2014,P15_XVI_InS,P18_VII_InS}. Following this approach, a local-variance map is built by computing the variance of those pixels in a map of resolution $N_{\mathrm{side}}$ that fall within discs of a given angular radius $R$, the centres of which are defined to be the pixel centroids of a lower resolution, $N_{\mathrm{side}}'$, map. The dipolar modulation is then characterised by fitting a dipole, $\vec{p}$, to the local-variance map, $d(\hat{n})$, from a weighted $\chi^2$,
\begin{equation}\label{eq_6_1}
    \chi^2 = \sum_{\hat{n}} \frac{[ (d(\hat{n})-\bar{d}(\hat{n})) - p_{0} - \hat{n}\cdot\vec{p}]^2}{\sigma^{2}(\hat{n})} \, ,
\end{equation}
where $\bar{d}(\hat{n})$ and $\sigma(\hat{n})$ are the mean and standard deviation of local-variance maps computed from random isotropic simulations, respectively. The best-fit value for the monopole ($p_{0}$) and dipole components ($p_x,p_y,p_z$) are calculated by minimising $-2\ln \mathcal{L} =\chi^2$. Note that the amplitude of this dipole, $P=|\vec{p}|$, is a positive definite quantity, and, despite being proportional to it, is not a direct measurement of the amplitude of the dipolar modulation.

We analyse the unconstrained and constrained simulation sets described in section~\ref{sec:simulations}, specifically those where the temperature data are inpainted on the inpainting mask.  
For this analysis, we work at resolutions $N_{\mathrm{side}}=64$ and $N_{\mathrm{side}}'=16$,  and adopt $R=6^\circ$ discs as a compromise between the higher sensitivity to the dipole amplitude seen for smaller radii in ref.~\cite{P18_VII_InS}, and having at least 100 pixels per disc to determine the variance effectively.

\begin{figure}
    \centering
    \includegraphics[width=\textwidth]{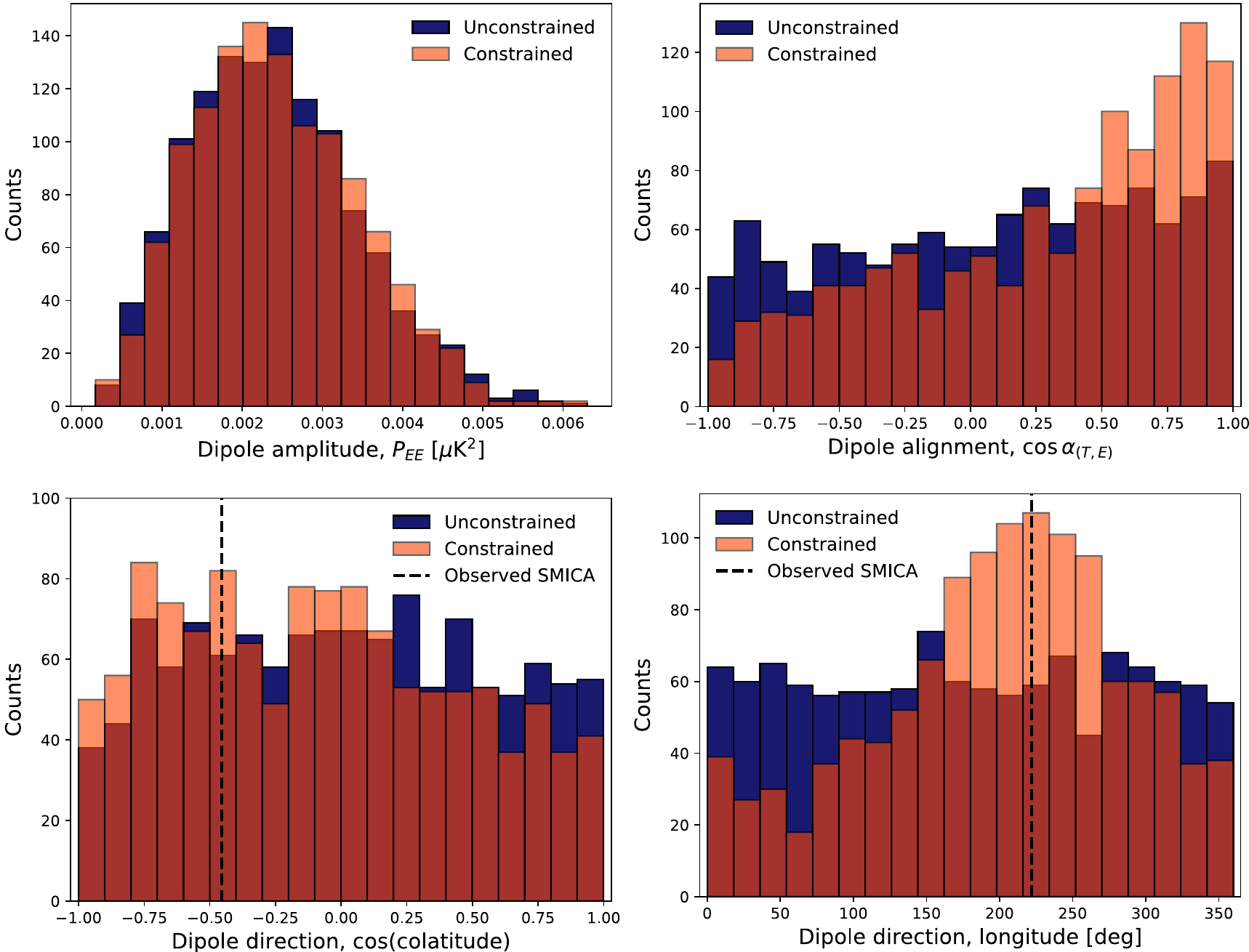}
    \caption{Histograms of the distributions of the dipole amplitude and direction obtained when fitting a dipole to maps of the local $E$-mode variance for unconstrained (blue) or constrained (orange) realisations, with the overlap shown in red. For each realisation, $\alpha_{(T,E)}$ is the angle between the dipoles found in the temperature and $E$-mode local-variance maps. In the bottom panels, the direction of the dipolar modulation found in the \texttt{SMICA} temperature map using the inpainting mask is indicated with a vertical black dashed line. }
    \label{fig:dipol_mod_EE}
\end{figure}

Initially, we consider the temperature-correlated part of the $E$-mode maps to determine whether the \smica-constrained realisations are anomalous. We find that, for a full sky analysis, the significance of the anomaly is diluted due to the inpainting effect. Specifically, if we define a {\it p}-value as the fraction of unconstrained realisations with dipole amplitudes equal to or larger than those observed in the constrained realisations,  we find a value of around 3.7\,\%.
Therefore, we repeat the analysis after applying the $E$-mode inpainting mask.
Note that the estimator is also slightly modified in this case such that only those discs in which more than 10\% of the selected pixels are unmasked are included when computing eq.~(\ref{eq_6_1}). The corresponding {\it p}-value is now reduced to 0.8\,\%.  We then proceed to analyse maps of the full $E$-mode signal. The results can be seen in figure~\ref{fig:dipol_mod_EE}, where $P_{EE}$ refers to the amplitude of the dipole in the local variance map of the $E$-modes.

The distributions of the dipole amplitude (top left panel) derived from the constrained and unconstrained realisations are essentially indiscernible. Therefore, the dipolar modulation found in the \smica\ map is not sufficient to induce a modulation of measurable amplitude in the $E$-mode polarisation signal. Nevertheless, in comparison with the random realisations, constrained simulations do show a modest excess of dipoles pointing roughly in the same direction as the modulation in the \texttt{SMICA} temperature data (bottom panels). This small alignment is better seen in the distribution of the cosine of the angle $\alpha_{(T,E)}$ between the dipoles found in the temperature and $E$-mode local-variance maps for each realisation (top right panel).

\begin{figure}
    \centering
    \includegraphics[width=\textwidth]{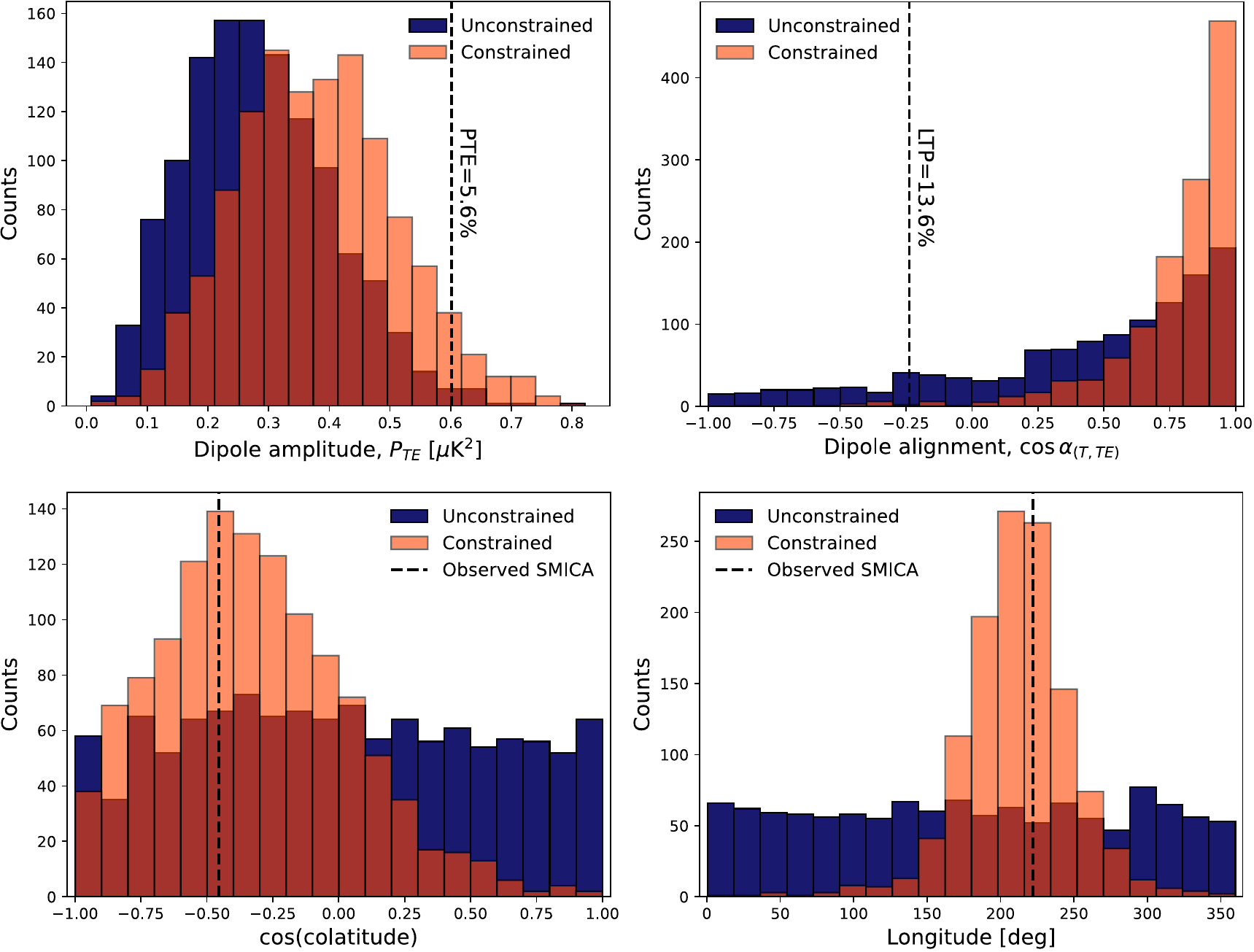}
    \caption{As in figure~\ref{fig:dipol_mod_EE}, but for the local $TE$ covariance.}
    \label{fig:dipol_mod_TE}
\end{figure}

A possible approach to enhance the sensitivity to very weak modulations would be to exploit the $TE$ correlation. We therefore repeat the previous analysis, and determine the best-fit dipoles to maps of the local $TE$ covariance. The results are shown in figure~\ref{fig:dipol_mod_TE}, where $P_{TE}$ refers to the amplitude of the dipole in the local covariance maps of the $T$ and $E$ modes.

The distribution of constrained realisations is now seen to peak at a slightly higher amplitude than that of the unconstrained realisations (top left panel), although the shift remains too small to claim a clear distinction between them. In particular, only 5.6\% of the unconstrained simulations are above the 99\% of the constrained ones. The dipoles fitted to the $TE$ local-covariance of the constrained realisations are clearly concentrated around the direction of the modulation in \texttt{SMICA} temperature (bottom panels). However, random realisations also indicate a substantial alignment between the dipoles found in temperature and in the $TE$ covariance (top right panel). Hence, finding a good alignment, $\cos \alpha_{(T,TE)} \approx 1$, between the preferred directions of the \texttt{SMICA} temperature and \LB\ $E$-mode polarisation maps is not sufficient to test the fluke hypothesis. 
However, we note that only 1\% of the constrained realisations yield $\cos \alpha_{(T, TE)} < -0.23$, whereas 13.6\% of the unconstrained simulations fall below this threshold.
Therefore, if $\cos \alpha_{(T,TE)} < -0.23$, then we can reject the fluke hypothesis at 99\% confidence level with a probability of 13.6\%. This defines the lower tail probability (LTP), indicated in the top right panel of the figure.

We conclude that, since the temperature dipolar modulation found in the \texttt{SMICA} data is insufficient to result in a statistically significant detection of modulation in the $E$-mode polarisation, any statistically significant dipolar modulation found therein by \LB, with an amplitude exceeding 0.005 $\mu \mathrm{K}^{2}$ would be independent evidence against the $\Lambda$CDM cosmological model, whereas significant anti-alignment of the dipolar modulation directions for temperature and $TE$ would argue against the fluke hypothesis.

\section{Low-$\ell$ alignment}
\label{sec:lowl_alignment}

\begin{figure}[t!]
    \centering
    \includegraphics[width=\textwidth]{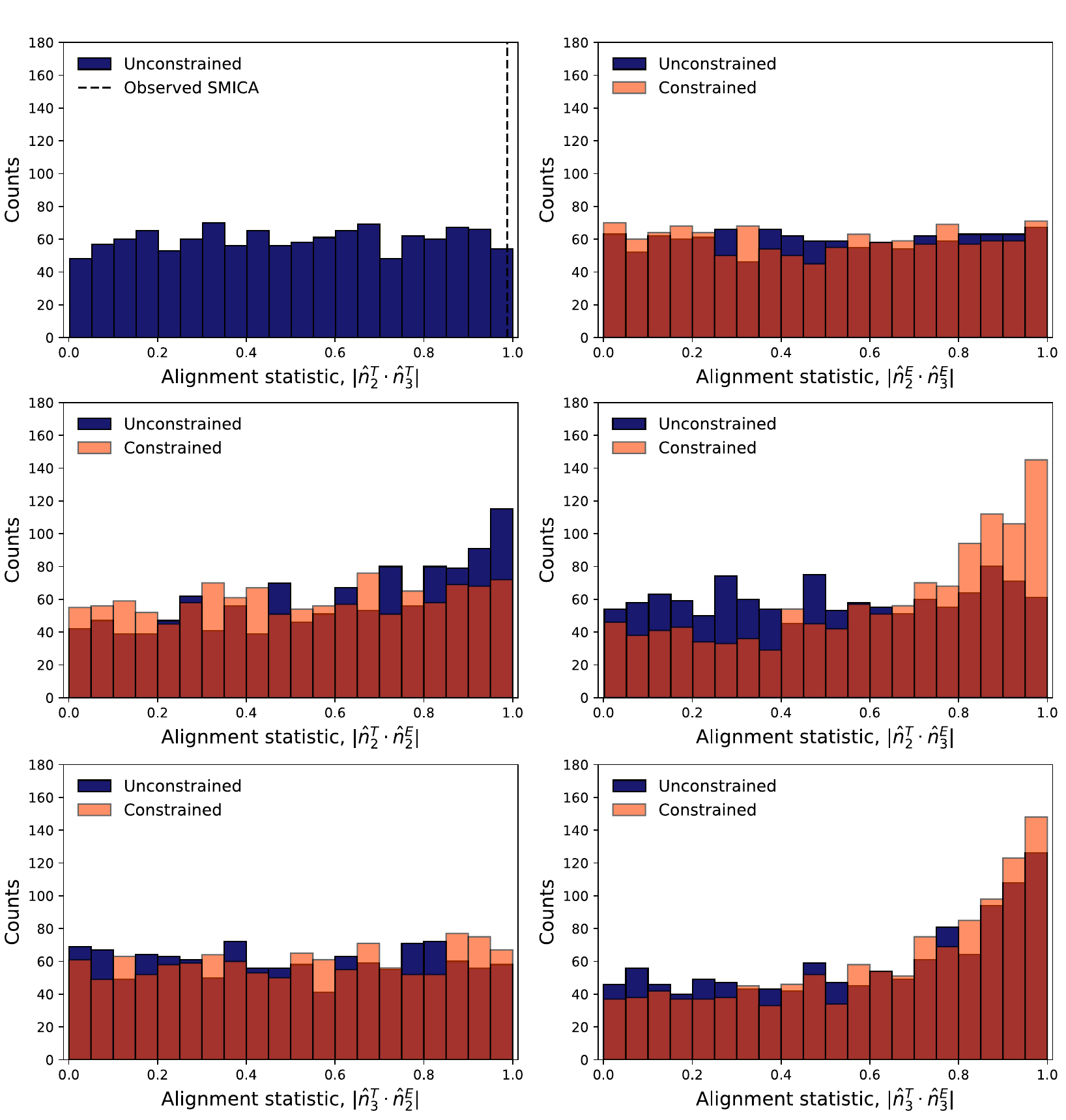}
    \caption{Alignment statistics computed between all possible pairs of preferred directions for temperature and $E$-mode maps at $\ell\,=\,2,3$. The histograms show the expected distributions of the statistics for 1200 unconstrained (blue) or constrained (orange) realisations, with the overlap of the distributions in red. Each of the $E$-mode maps is constrained by a separate realisation of the \Planck\ \texttt{SMICA} temperature map inpainted using the inpainting mask. The top left panel indicates the temperature alignment statistic, where the observed value is determined from the mean of the values computed from the inpainted \texttt{SMICA} maps and indicated by a black dashed line.}
     \label{fig:lowl_alignment}
\end{figure}

A significant alignment between the orientation of the quadrupole ($\ell=2$) and octopole ($\ell=3$) of the \wmap\ first year temperature data, unexpected if the CMB is an isotropic random field, was first noted by refs.~\cite{Tegmark2003,deOliveiraCosta2004,Schwarz2004}. The presence of this feature of the CMB sky in the \Planck\ first year data was later confirmed in refs.~\cite{P13_XXIII_InS,Copi:2015}.

By considering the spherical harmonic coefficients, $a^{X}_{\ell m}(\hat{n})$, of an $X$-mode CMB map, where $X$ corresponds to either $T$, $E$ or $B$, in a rotated coordinate system such that the $z$-axis lies in the $\hat{n}$-direction, ref.~\cite{deOliveiraCosta2004} determines a preferred axis when maximising the angular momentum dispersion for a given $\ell$
\begin{equation}
\sum_m m^2 |a^{X}_{\ell m}(\hat{n})|^2 .
\label{eq:ang_mom_disp}
\end{equation} 
Note that this quantity does not distinguish between $\hat{n}$ and $-\hat{n}$.
The angular separation between the preferred axes for the temperature quadrupole, 
$\hat{n}^T_2$, and octopole, $\hat{n}^T_3$, is then determined from the quantity 
$| \hat{n}^T_2 \cdot \hat{n}^T_3 |$. We refer to this as the alignment statistic.

Here, we compare the alignment statistics for all permutations of the temperature and $E$-mode quadrupole and octopole data drawn from two sets of 1200 unconstrained or constrained simulations, as described in section~\ref{sec:simulations}. 
In order to examine whether the inpainting process biases the results of a full sky analysis, 
we test whether each of the 1200 inpainted \smica\ temperature realisations are anomalous by defining a {\it p}-value with reference to the inpainted unconstrained simulations. This is defined as the fraction of unconstrained simulations with an alignment statistic (angular separation) at least as large (small) as the inpainted \smica\ realisation. For data inpainted on the inpainting mask, all realisations return {\it p}-values of $0.01$ or below, with a median of $0.008$.  
In the case of the common mask, the median remains at $0.008$ but with a
larger scatter, and increasing to a maximum of around 0.1. However, $99\,\%$ of the realisations return {\it p}-values below 0.05. We conclude that the significance of the anomaly is maintained when performing a full-sky analysis of the inpainted maps. Then, for this particular statistic, we will use maps inpainted on the inpainting mask due to the lower scatter of their {\it p}-values.
Since the $\ell\, = 2$ and $\ell\, = 3$ temperature-correlated $E$-mode spherical harmonics are derived by a simple scaling of the corresponding temperature modes, their preferred directions and alignment statistics are identical and also anomalous.

Figure~\ref{fig:lowl_alignment} presents the results from our analysis of the realisations where the temperature maps are inpainted using the inpainting mask. In the top left panel, the blue histogram shows the expected behaviour for the unconstrained temperature simulations. Since the quadrupole and octopole are statistically independent quantities for an isotropic random field, the alignment statistic follows a uniform distribution on the unit interval. The black dashed line corresponds to the median value of $0.988$ computed from all of the inpainted \smica\ realisations, a misalignment between the quadrupole and octopole of $9^{\circ}$,
with a {\it p}-value of $0.008$. This is in good agreement with the values reported in ref.~\cite{P13_XXIII_InS}. The top right panel shows the equivalent result for the $E$ modes.
Both the unconstrained and constrained distributions seem to be consistent with a uniform distribution. This might be considered unexpected in the latter case, since the temperature-correlated parts of the $E$-mode quadrupole and octopole must reflect the alignment of the \smica\ temperature modes. However, it has long been observed that the CMB quadrupole amplitude in temperature is unusually low \cite{Hinshaw:1996bps,WMAP:2003ivt}, so that the part of the $E$-mode quadrupole signal that is uncorrelated with the temperature quadrupole dominates the variance of the statistical distribution, and the behaviour of the alignment statistic approximates an isotropic field. 

The middle panels show the alignment statistic computed between the temperature quadrupole and $E$-mode quadrupole (left) and octopole (right). In the case of $| \hat{n}^T_2 \cdot \hat{n}^E_2 |$, the unconstrained simulations show a modest preference for alignment that reflects the intrinsic correlation between the temperature and $E$-mode quadrupoles. However, this behaviour is not seen for the constrained realisations, which is again related to the low amplitude of the \smica\ quadrupole. For $| \hat{n}^T_2 \cdot \hat{n}^E_3 |$, the unconstrained simulations are consistent with statistically independent quantities, whereas the constrained realisations do show a modest preference for alignment, due to the alignment between the \smica\ temperature quadrupole and \smica-constrained $E$-mode octopole. The \smica\ temperature octopole has an amplitude consistent with that expected for the \Planck\ 2018 cosmological model.

The bottom panels present the results for the temperature octopole and $E$-mode quadrupole (left) and octopole (right) alignment statistics. In the former case, the distributions for both unconstrained and constrained realisations are consistent with that expected for statistically independent quantities, where the potential alignment in the constrained simulations are again suppressed by the effect of the low \smica\ quadrupole. In the latter case, both sets of simulations show a modest preference for alignment as expected. It should be apparent that there are no values of the alignment statistics presented here that can provide insight into the fluke hypothesis.

\section{Harmonic-based estimators}
\label{sec:harmonic}

In this section, we consider harmonic-based estimators designed to study anomalies detected in the CMB temperature maps of \WMAP\ \cite{Bennett2011} and \Planck\ data \cite{P13_XXIII_InS,P15_XVI_InS,P18_VII_InS} at low multipoles. In particular, we focus on the point-parity or even-odd asymmetry \cite{Land:2005jq,Kim:2010gf,Kim:2010gd,Gruppuso:2010nd}, as well as the lack of signal on large angular scales in the two-point angular correlation function \cite{Hinshaw:1996ut,WMAP:2003elm,Bernui:2006ft,Copi:2006tu,Copi:2008hw,Efstathiou:2009di,Gruppuso:2013dba}.
These features are clearly present in the CMB temperature data; however, their statistical significance is generally low, around 2-3$\,\sigma$. Therefore, incorporating the information contained in the polarised CMB anisotropy pattern may help to determine whether these features are merely statistical flukes or traces of deviations from the standard $\Lambda$CDM cosmological model.
For this reason, we use the information contained in the $EE$ and $TE$ CMB angular power spectra and build estimators for the even-odd asymmetry in section~\ref{sec:evenoddest}, and the lack-of-correlation anomaly in section~\ref{sec:lack_of_corr}.
To construct these harmonic-based estimators, we first compute the CMB power spectra from the constrained and unconstrained realisations described in section~\ref{sec:simulations}. This computation was performed using \texttt{ECLIPSE} \cite{Bilbao-Ahedo:2021jhn}, an implementation of a quadratic maximum likelihood estimator \cite{Tegmark:2001zv}. The analysis was conducted at \nside\ = $64$, employing the common and the inpainting mask for temperature, while applying the corresponding extended polarisation masks derived in section~\ref{sec:data} to the $Q$ and $U$ sky maps.

\subsection{Even-odd asymmetry}
\label{sec:evenoddest}

An estimator for the point-parity symmetry of the large angular-scale anisotropy of the CMB 
was defined in ref.~\cite{Land:2005jq} based on the properties of the even and odd $C_{\ell}$ power spectra, and proposed as a practical tool for detecting foreground residuals. However, contrary to expectations, they found that the \WMAP\ data exhibited an odd-parity preference, though only at a modest $95\%$ confidence level. Later, ref.~\cite{Kim:2010gf} identified an anomaly in the parity symmetry of the \WMAP\ 3-year and 5-year temperature maps, with a significance level of 4 in 1000 at large angular scales. This finding was later confirmed in the \WMAP\ 7-year data in refs.~\cite{Kim:2010gd,Gruppuso:2010nd,Aluri:2011wv} at a similar level of significance. These analyses were further extended to \Planck\ data from the 2013 \cite{P13_XXIII_InS}, 2015 \cite{P15_XVI_InS}, and 2018 \cite{P18_VII_InS} data releases, consistently finding an anomaly at the percent level or below, with a maximum significance reaching 0.2-0.4\,\%. In the latter case, it was shown that, when accounting for the look-elsewhere effect, such an anomaly decreases to approximately 1.6-2\,\%.

\begin{figure}[t!]
    \centering
\includegraphics[width=1.0\textwidth]{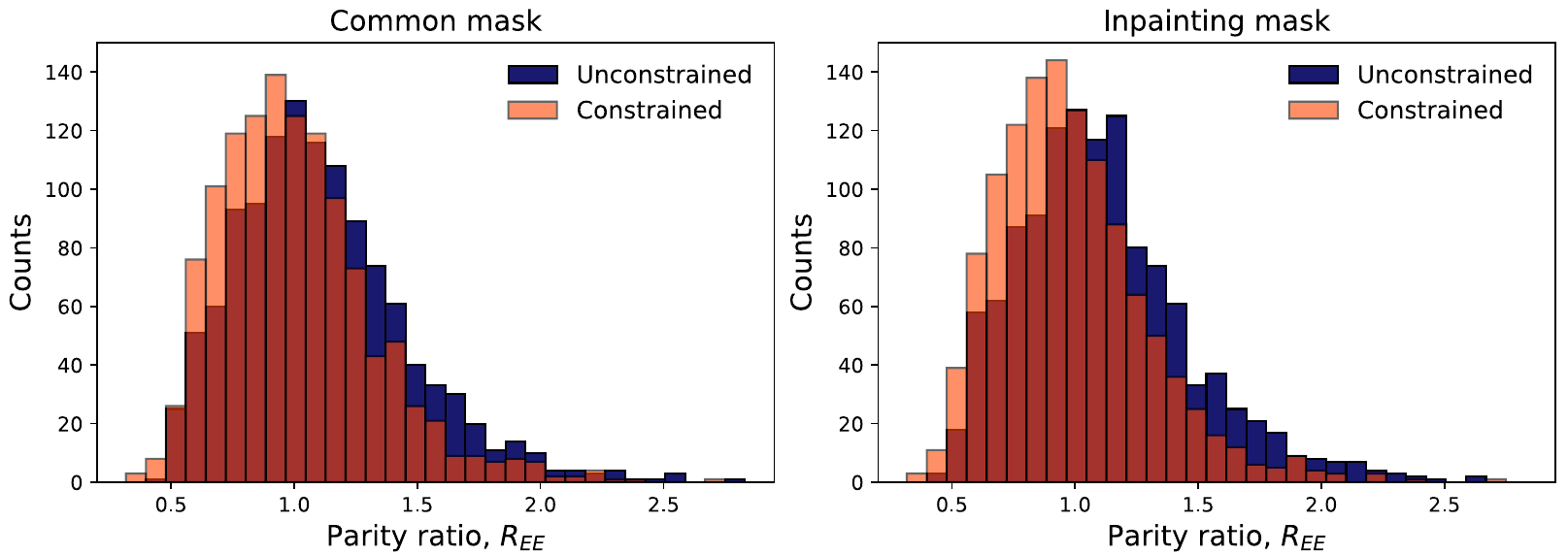}
    \caption{Histograms of $R_{EE}(\ell_{\text{max}})$ computed with $\ell_{\text{max}}=24$ from the unconstrained simulations (blue) and the constrained simulations (orange), with the overlap shown in red. Similar plots are obtained for other values of $\ell_{\text{max}}$ from 3 to 50. The left panel presents results derived using the common mask, and the right panel for the inpainted mask.}
    \label{fig:REE}
\end{figure}

By generalising to $E$ modes the estimator used to study the even-odd asymmetry \cite{Kim:2010gd,Kim:2010gf} in temperature, one can write
\begin{equation}
    R_{EE}(\ell_{\text{max}}) = \frac{C^{+}_{EE}(\ell_{\text{max}})}{C^{-}_{EE}(\ell_{\text{max}})} \, ,
    \label{RXX}
\end{equation}
where
\begin{equation}
    C^{+/-}_{EE}(\ell_{\text{max}}) = \frac{1}{N_{+/-}} \, \sum_{\ell \in \, \text{even/odd}}^{\ell_{\text{max}}} \frac{\ell (\ell+1)}{2 \pi} C_{\ell}^{EE} \, ,
    \label{Cplusminus}
\end{equation}
with $N_{+/-}$ being the number of even/odd multipoles present in the considered harmonic range (i.e., from 2 to $\ell_{\text{max}}$), where the sum is taken over the even/odd multipoles. 
However, it can be shown that $R_{EE}$ is unable to rule out the fluke hypothesis, since the histograms constructed for the constrained and unconstrained total $E$-mode signal are practically indistinguishable (see figure~\ref{fig:REE} for $R_{EE}(\ell_{\text{max}}=24)$, where the left panel corresponds to the common mask and the right panel to the inpainted mask). The choice of $\ell_\mathrm{max} = 24$ serves as a representative example of the behaviour of this estimator, with very similar results observed for other values of $\ell_{\rm max}$.

\begin{figure}[t!]
    \centering
    \includegraphics[width=0.95\textwidth]{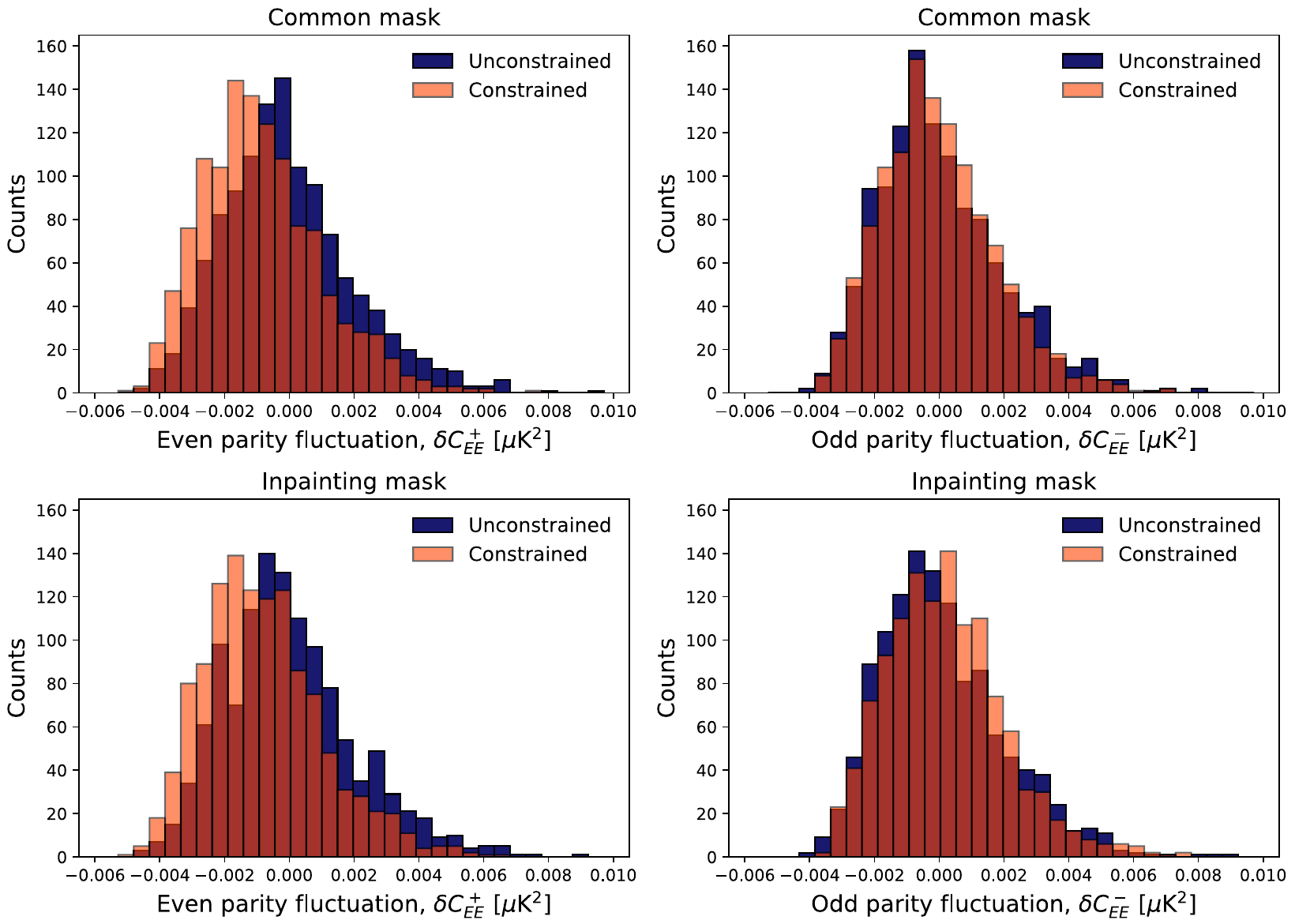}
    \caption{Histograms of $\delta C^{+}_{EE}$ (left panels) and $\delta C^{-}_{EE}$ (right panels) computed with $\ell_\mathrm{max}=27$ from the unconstrained (blue) and constrained simulations (orange), with the overlap in the distributions shown in red. The upper panels correspond to results for the common mask, while the lower panels are for the inpainting mask.}
    \label{fig:deltaCEE}
\end{figure}

We then consider an analysis based on the $TE$ spectrum. Although this is not connected to the point-parity of a map, it is not unreasonable to study the relative behaviour of the even and odd modes of the spectrum.
In this case, we observe a noticeable shift between the histograms of $R_{TE}(\ell_{\text{max}})$ constructed from the constrained and unconstrained simulations, where $R_{TE}$ is built as in eq.~(\ref{RXX}) with $EE$ replaced by $TE$. For $\ell_{\text{max}} = 27$, this shift is most pronounced in the case of the common mask, where the probability of ruling out the fluke hypothesis is approximately $24.6\%$, while the average probability for $\ell_{\text{max}} \in [17,31]$ is $19.9\%$. Similarly, for the inpainting mask, the probability peaks at $30.5\%$ for $\ell_{\text{max}} = 22$, with an average of $27.5\%$ over the range $\ell_{\text{max}} \in [19,30]$. These probabilities are computed by evaluating the area of the histogram of the open simulations to the right of the 99th percentile value of the constrained distribution. 
Note that, although the $TE$ spectrum is on average positive within the harmonic considered range, some realisations, particularly at low $\ell$, may take on negative values. This can result in a negative $R_{TE}$ estimator, especially at very low $\ell_{\text{max}}$, when the denominator becomes much smaller than the numerator, approaching zero from below. Although this behavior is traced by the simulations, it suggests that $R_{TE}$ may not be ideal for this type of analysis, particularly at low $\ell_{\text{max}}$.
For the above reasons, we also consider an alternative approach based on the analysis of the even and odd $C_{\ell}^{XX}$ separately, see e.g.~ref.~\cite{Gruppuso:2017nap}. This can be achieved by considering the mean fluctuations of the even or odd multipoles with respect to the fiducial spectrum $C_{\ell}^{XX,\text{fid}}$, i.e.~$\delta C^{+/-}_{XX}$, where $XX$ denotes $TT$, $EE$, or $TE$, defined as follows,
\begin{equation}
    \delta C^{+/-}_{XX}(\ell_\mathrm{max}) = \frac{1}{N_{+/-}} \, \sum_{\ell \in \, \text{even/odd}}^{\ell_\mathrm{max}} \frac{\ell (\ell+1)}{2 \pi} (C_{\ell}^{XX}-C_{\ell}^{XX,\text{fid}}) \, ,
    \label{deltaCX}
\end{equation}
where the factor $\ell(\ell+1)/2 \pi$ ensures that each term in the sum has (almost) equal weight in $TT$.
In appendix~\ref{app:evenodd}, we show that $\delta C^{+}_{TT}$ is anomalous in the \Planck\ data with $\ell_\mathrm{max} \in [20,30]$, while $\delta C^{-}_{TT}$ is compatible with the $\Lambda$CDM expectation.\footnote{Therefore in this approach the even-odd asymmetry is interpreted as a lack of power in the even multipoles.} Moreover, in the same appendix we verify that $\delta C^{+}_{EE}$ is able to detect the anomaly with a significance comparable to that found in $TT$, if built with the temperature-correlated $E$-mode data. 
We then compute the $\delta C^{+}_{EE}$ and $\delta C^{-}_{EE}$ statistics for the full $E$-mode signal, as shown in the left and right panels of figure~\ref{fig:deltaCEE}. The upper panels correspond to results derived with the common mask, while the lower panels correspond to the inpainting mask.
The histograms are derived from both the unconstrained simulations (in blue) and the constrained simulations (in orange), with $\ell_{\text{max}} = 27$.
Unfortunately, it turns out that eq.~(\ref{deltaCX}) with $X=EE$ is insufficient to construct a suitable estimator for the even-odd asymmetry, since the difference between the constrained and unconstrained histograms is minimal. The behaviour of $\delta C^{+/-}_{EE}$ shown in figure~\ref{fig:deltaCEE} for $\ell_{\text{max}} = 27$ remains consistent across the range of $\ell_{\text{max}}$ values considered (from 3 to 50).

\begin{figure}[t!]
    \centering    
    \includegraphics[width=0.95\textwidth]{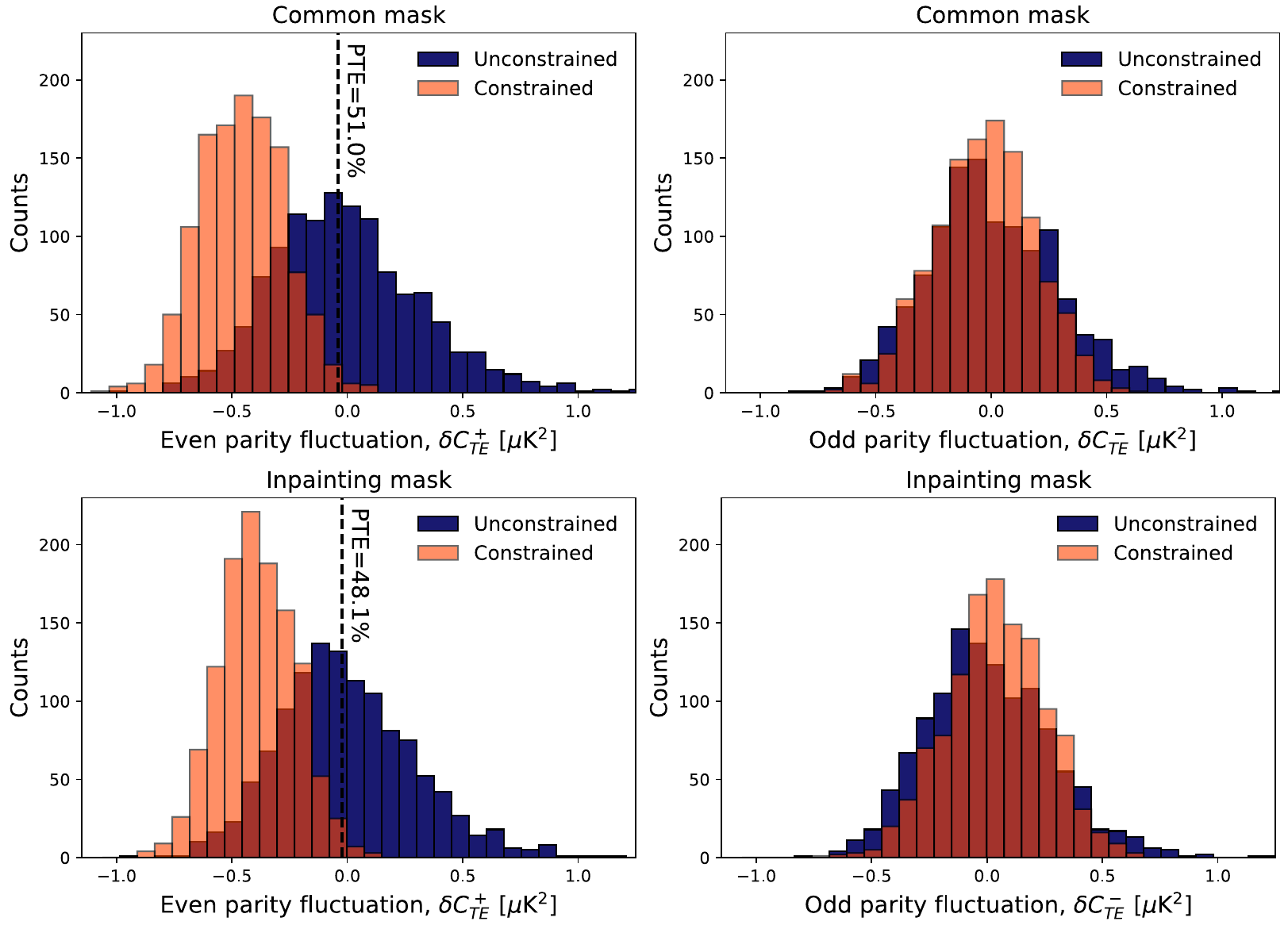}
    \caption{Histograms of $\delta C^{+}_{TE}$ (left panels) and $\delta C^{-}_{TE}$ (right panel), computed from both the unconstrained (blue) and constrained (orange) simulations, with the overlap between the distributions shown in red. The upper panels correspond to the common mask and refer to $\ell_{\text{max}} = 25$, while the lower panels correspond to the inpainting mask and refer to $\ell_{\text{max}} = 27$.}
    \label{fig:CTE}
\end{figure}

\begin{figure}[t!]
    \centering    
    \includegraphics[width=0.6\textwidth]{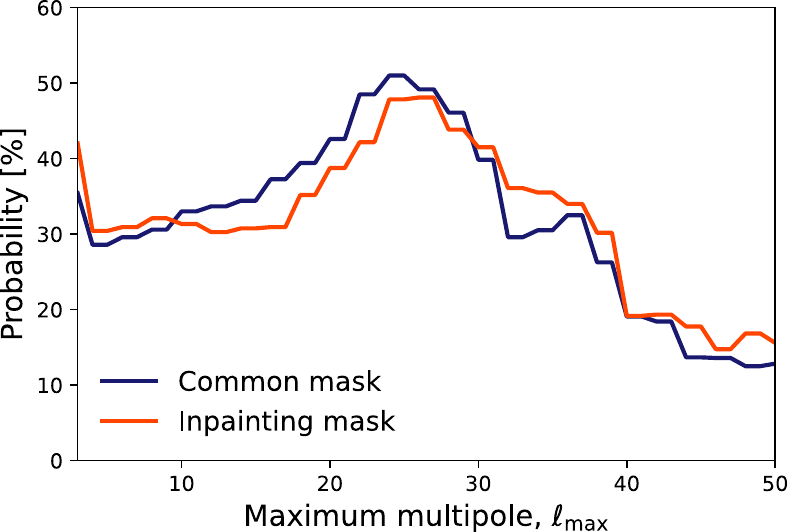}
    \caption{Probability of ruling out the fluke hypothesis versus $\ell_{\rm max}$ considering $\delta C^{+}_{TE}$ for the common mask (blue line) and the inpainting mask (orange line).}
    \label{fig:CTE+percentage}
\end{figure}

Finally, we consider the estimators $\delta C^{+/-}_{TE}$.
Figure~\ref{fig:CTE} presents the results for $\delta C^{+/-}_{TE}$ derived from the constrained and unconstrained simulations with $\ell_{\text{max}} = 25$ for the common mask (upper panels) and with $\ell_{\text{max}} = 27$ for the inpainting mask (lower panels).
Moreover, the left panels are for $\delta C^{+}_{TE}$ and the right panels for $\delta C^{-}_{TE}$
While the right panels do not show any displacement between the two histograms, a clear shift is instead observed in the left panels which is based on even multipoles.
Similar histograms are obtained for $\ell_{\text{max}}$ values ranging from 3 to 50, where $\delta C^{-}_{TE}$ does not exhibit any measurable shift between the two histograms, while the amplitude of the shift in $\delta C^{+}_{TE}$ varies.
The highest probability of rejecting the fluke hypothesis is found to be $51.0\%$ for the common mask at $\ell_{\text{max}} = 25$ whereas it is $48.1\%$ for the inpainting mask at $\ell_{\text{max}} = 27$. These probabilities are computed by evaluating the right tail of the histogram obtained from the unconstrained simulations, starting from the value corresponding to the 99th percentile of the constrained simulations.
Figure~\ref{fig:CTE+percentage} shows the probability of rejecting the fluke hypothesis as a function of $\ell_{\text{max}}$, based on $\delta C^{+}_{TE}$ (blue line for the common mask and orange line for the inpainting mask). 

\subsection{Lack of correlation}
\label{sec:lack_of_corr}
The measured two-point angular correlation function of the CMB temperature anisotropies has been found to exhibit a lack of large-angle correlations as compared to the standard cosmological model.
Initially observed in the \textit{COBE-DMR} data in the 1990s \cite{Hinshaw:1996ut,Bennett:1996ce}, subsequent studies based on the \wmap\ data
characterised the anomaly as a surprising deficit in correlation power on angular scales larger than $60^\circ$ \cite{WMAP:2003ivt,Copi:2006tu}. 
This effect has persisted through multiple data releases of the \planck\ mission \cite{P18_VII_InS}, remaining robust across various analysis methods, sky masks, and component-separation techniques.

To quantify this deficit, the \wmap\ team defined the $S_{1/2}$ statistic \cite{WMAP:2003elm} as the integral of the square of the two-point angular correlation function $C(\theta)$ over the angular range from $60^\circ$ to $180^\circ$. This statistic was later generalised in ref.~\cite{Copi:2013zja} as the $S^{XY}$ statistic, which measures correlations between any combination of temperature ($T$) and polarisation ($Q$, $U$) maps; see also ref.~\cite{Chiocchetta:2020ylv} for a recent analysis on \Planck\ data. While the observed value of $S^{TT}$ for temperature correlations has consistently been shown to be low, with {\it p}-values typically less than 1\% for $\Lambda$CDM simulations, polarisation data may provide a crucial test to distinguish between a mere statistical fluke and genuine new physics. If the lack of large-angle temperature correlations is simply a statistical fluctuation within the $\Lambda$CDM framework, then the temperature-polarisation cross-correlation should also exhibit suppression at large angles, at a level consistent with the intrinsic correlation between the temperature and $E$-mode polarisation signals. In this section, we present an analysis to address the fluke hypothesis and explore its implications. 

The estimator which quantifies the distance of the two-point angular correlation function from zero over the range $[\theta_1, \theta_2]$ is given by
\begin{equation}
    S^{XY}(\theta_1, \theta_2) \equiv \int_{\cos\theta_2}^{\cos\theta_1} \,  \left[ C^{XY}(\theta)\right]^2 \, d(\cos \theta)
    \, ,
    \label{eq:SXX}
\end{equation}
where $C^{XY}(\theta)$ is the two-point correlation function of the $XY$ spectrum.
Often, we set $\theta_1=60^{\circ}$ and $\theta_2=180^{\circ}$, consistent with the earlier \wmap\ analysis, and refer to the statistic in this case as $S^{XY}_{1/2}$. 
Here, we study $S^{XY}$ estimators based on $E$ modes (section \ref{sec:lack_of_corr_EE}), $TQ$ (section \ref{sec:lack_of_corr_TQ}) and $TE$ (section \ref{sec:lack_of_corr_TE}).

\subsubsection{Estimators based on $E$ modes}
\label{sec:lack_of_corr_EE}
We focus now on $XY=EE$. The two-point correlation function, $C^{EE}(\theta)$, is defined as
\begin{equation}
   C^{EE}(\theta) = \sum_{\ell=2}^{\ell_\mathrm{max}} \frac{2 \ell +1}{4 \pi} \, C_{\ell}^{EE} P_{\ell}(\cos \theta)\, , \end{equation}
where the $P_{\ell}(x)$ are the Legendre polynomials.
Then, following ref.~\cite{Copi:2008hw}, eq.~(\ref{eq:SXX}) can be written as
\begin{equation}
    S^{EE}_{1/2} = \sum_{\ell=2}^{\ell_\mathrm{max}} \sum_{\ell^{\prime}=2}^{\ell_\mathrm{max}} \, \frac{2 \ell +1}{4 \pi} \, \frac{2 \ell^{\prime} +1}{4 \pi} \,
    C_{\ell}^{EE} \, I_{\ell \ell^{\prime}} \, C_{\ell^{\prime}}^{EE} \, ,
    \label{S1/2EE}
\end{equation}
where
\begin{equation}
    I_{\ell \ell^{\prime}} = \int_{-1}^{1/2} \,P_{\ell}(x) P_{\ell^{\prime}}(x) \, dx \, .
\end{equation}
Note that $I_{\ell \ell^{\prime}}$ is computed following appendix~A of ref.~\cite{Copi:2008hw}.

\begin{figure}[t!]
    \centering
    \includegraphics[width=\textwidth]{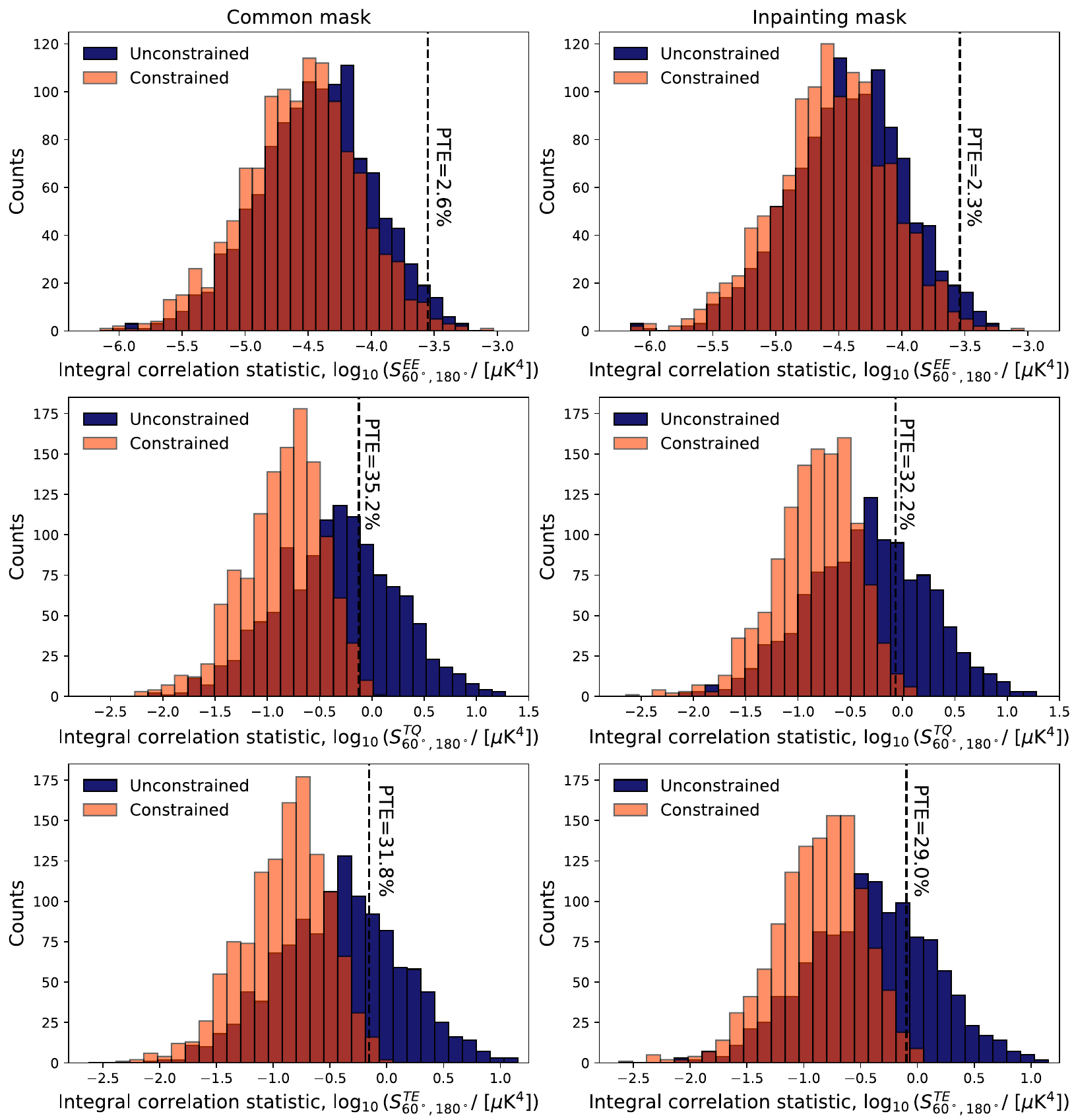}
    \caption{Histogram of $S^{EE}_{1/2}$ (top row), 
    $S^{TQ}_{1/2}$ (middle row), and $S^{TE}_{1/2}$ (bottom row), 
    for the unconstrained simulations (blue) and the constrained simulations (orange), with the overlap of the distributions shown in red, for $\ell_\mathrm{max}=$ 32. The dashed
    vertical line represents the threshold used to calculate the 99\%
    PTE, whose value is reported in the legend.  
    The results computed for the common mask are shown in the left column, and those for the inpainting mask in the right column.}
    \label{fig:S1/2_EETQTE}
\end{figure}

We then compute histograms from the $S^{EE}_{1/2}$ values determined from the constrained and unconstrained simulations described in section~\ref{sec:simulations}, for the common and inpainting masks presented in section~\ref{sec:data}. 
Here, we report on results derived for $\ell_\mathrm{max}=32$, since these are largely insensitive to variations in $\ell_\mathrm{max}$ across the range of 10 to 50. We then calculate the 99\% PTE from the fraction of unconstrained simulations that yield a value of $S^{EE}_{1/2}$ above a given threshold, defined as the value of $S^{EE}_{1/2}$ such that 99\%  of the constrained simulations have lower values. The results are presented in the top row of figure~\ref{fig:S1/2_EETQTE}, with the threshold shown as a dashed 
vertical line, together with the corresponding PTE value. This can be interpreted as the probability of obtaining a realisation which is not compatible with that expected from the \texttt{SMICA} temperature observations. 
The low PTE values (up to 2.5\%) indicate considerable overlap of the two histograms computed from the constrained and unconstrained simulations, implying that the $S^{EE}_{1/2}$ estimator is not suitable for testing the fluke hypothesis.

\begin{figure}[t!]
    \centering
    \includegraphics[width=1.0\textwidth]{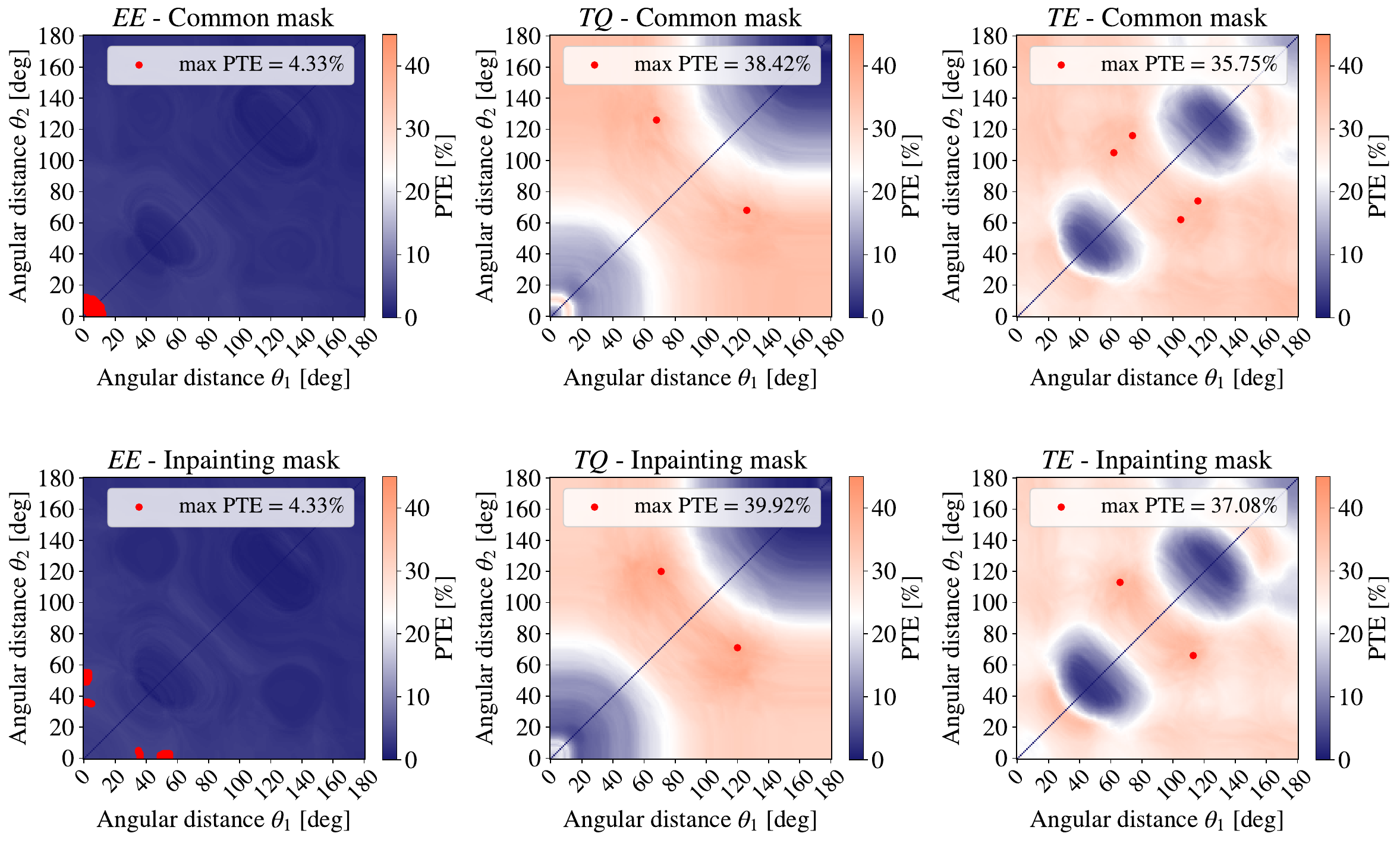}
    \caption{99\% PTE values for different $(\theta_1, \theta_2)$ angles for the $S^{EE}(\theta_1, \theta_2)$ (left panels), $S^{TQ}(\theta_1, \theta_2)$ (middle panels) and $S^{TE}(\theta_1, \theta_2)$ (right panels) estimators with $\ell_\mathrm{max}=$ 32, determined from the common mask (top row) and inpainting mask (bottom row) simulations. The red circles indicate the maximum PTE value within each matrix. The location of the maximum PTE value is consistent across both masks. However, for the $S^{EE}$ estimator, the situation is more nuanced, likely because this estimator is not well-suited to test the anomaly. Interestingly, for the $S^{TE}$ estimator using the common mask (top right panel), more than one point was found to have the maximum PTE value.}
    \label{fig:PTE_matrices_99}
\end{figure}

Subsequently, we seek an optimised $S^{EE}(\theta_1, \theta_2)$ statistic by considering a grid of angles for $\theta_{1,2}\in [0^{\circ},180^\circ]$ in steps of $1^\circ$. We then compute the 99\% PTE value for each $(\theta_1, \theta_2)$ pair and present the results in the form of a matrix in the top (common mask) and bottom (inpainting mask) left panels of figure~\ref{fig:PTE_matrices_99}. The maximum PTE value is highlighted in the matrix with red circles, since multiple pairs of angles yield the same maximum.  We report in table~\ref{tab:SXX_max_PTE} the 99\% maximum PTE values determined from simulations that adopt the $E$-mode common or inpainting masks, together with one pair of $(\theta_1, \theta_2)$ angles that maximises the constraining power of the $S^{EE}$ estimator. We find a maximum PTE value of 4.3\% for both the common and inpainting masks, occurring at relatively small angular scales of ($8^\circ,9^\circ$) and ($5^\circ,35^\circ$), respectively. 
We conclude that, even after a broad exploration of the behaviour of the statistic over the $(\theta_1, \theta_2)$ parameter space, the maximum PTE value remains low and the estimator is not suitable for testing the fluke hypothesis.

\subsubsection{Estimators based on $TQ$}
\label{sec:lack_of_corr_TQ}
We now perform the analysis based on the 2-point cross-correlation function between $T$ and $Q$,
\begin{equation}
    C^{TQ}(\theta) = \sum_{\ell=2} \frac{2\ell+1}{4\pi}\sqrt{\frac{(\ell-2)!}{(\ell+2)!}}C^{TE}_{\ell} P^2_{\ell}(\cos{\theta}) \, , 
\end{equation}
where $P^2_{\ell}(\cos{\theta})$ are the associated Legendre polynomials of order 2. 
Then following ref.~\cite{Copi:2013zja},
\begin{equation}
    S^{TQ}_{1/2} = \sum_{\ell = 2}^{\ell_{max}} \sum_{\ell' = 2}^{\ell_{max}}\frac{2\ell+1}{4\pi} \frac{2\ell'+1}{4\pi}\sqrt{\frac{(\ell-2)!}{(\ell+2)!}\frac{(\ell'-2)!}{(\ell'+2)!}}C^{TE}_{\ell}I^{TQ}_{\ell,\ell'}C^{TE}_{\ell'},
    \label{eq:shalfTQ}
\end{equation}
with
\begin{equation}
I^{TQ}_{\ell \ell'} \equiv \int_{-1}^{1/2} P^{2}_{\ell}(x)P^{2}_{\ell'}(x)dx \, ,
\end{equation}
where $I^{TQ}_{\ell \ell'}$ is computed following appendix~B of ref.~\cite{Copi:2013zja}.

The results of the analysis for these quantities are presented in the second row of figure~\ref{fig:S1/2_EETQTE}. In contrast to the $S^{EE}_{1/2}$ estimator, the histograms corresponding to the constrained and unconstrained simulations exhibit considerably reduced overlap. The PTE is thus larger, meaning that there is a larger probability to obtain a realisation which is not compatible with expectations when constrained by temperature observations. Therefore, $S^{TQ}_{1/2}$ is an improved estimator for testing the fluke hypothesis and should be considered for future studies of the lack of correlation anomaly.

Results from the optimisation study for the 99\% PTE are shown in the top (common mask) and bottom (inpainting mask) middle panels of figure~\ref{fig:PTE_matrices_99} and summarised in table~\ref{tab:SXX_max_PTE}. The maximum PTE value is observed on the $E$-mode inpainting mask, reaching 39.9\% ($71^\circ,120^\circ$). However, the optimisation study consistently reveals higher PTEs for the $E$-mode common mask, as illustrated in figure~\ref{fig:PTE_matrices_99}. It is also worth noting that, although the PTE is only slightly higher than in the standard ($60^\circ,180^\circ$) case, the maximum values for both masks occur at a similar location in the parameter space and correspond to large angular scales, in contrast to the $EE$ case.

\subsubsection{Estimators based on $TE$}
\label{sec:lack_of_corr_TE}
We finally consider the estimator based on the cross-correlation between $T$ and $E$:
\begin{equation}
    C^{TE}(\theta) = \sum_{\ell=2}^{\ell_{max}} \frac{2\ell+1}{4\pi}C^{TE}_{\ell} P_{\ell}(\cos{\theta}) \, . 
\end{equation}
Equation~(\ref{eq:SXX}) with $XX=TE$ then becomes
\begin{equation}
    S^{TE}_{1/2} = \sum_{\ell = 2}^{\ell_{max}} \sum_{\ell' = 2}^{\ell_{max}}\frac{2\ell+1}{4\pi} \frac{2\ell'+1}{4\pi}C^{TE}_{\ell}I_{\ell,\ell'}C^{TE}_{\ell'}.
    \label{eq:shalfTE}
\end{equation}

Results from this analysis are presented in the third row of figure~\ref{fig:S1/2_EETQTE}. As for the $S^{TQ}_{1/2}$ estimator, the histograms corresponding to the constrained and unconstrained simulations do not fully overlap, and the PTE is higher compared to that of the $S^{EE}_{1/2}$ estimator. The $S^{TE}_{1/2}$ estimator can also be considered suitable for testing the fluke hypothesis.

As before, we attempt to optimise the statistic for the 99\% PTE, shown in the top (common mask) and bottom (inpainting mask) right panels of figure~\ref{fig:PTE_matrices_99}, and summarised in table~\ref{tab:SXX_max_PTE}.
Similar to the $TQ$ case, the maximum PTE value occurs for the $E$-mode inpainting mask, reaching 37.1\% ($66^\circ,113^\circ$). However, the optimisation study consistently reveals larger PTEs for the $E$-mode common mask, as shown in figure~\ref{fig:PTE_matrices_99}. It is also worth noting that, similar to the $TQ$ case, the maximum PTE values occur at large angular scales, in contrast to the $EE$ case, and are located in a similar region of the parameter space for both masks.

\begin{table}
\begin{center}
\begin{tabular}{cccc}
\toprule
99\% PTE & $S^{EE}$ & $S^{TQ}$ &  $S^{TE}$  \\
\midrule
\multirow{2}{*}{Common} & 4.3\% & 38.4\% & 35.8\% \\
                        & $(8^\circ, 9^\circ)$ & $(68^\circ, 126^\circ)$ & $(62^\circ, 105^\circ)$ \\
\midrule
\multirow{2}{*}{Inpainting} & 4.3\% & 39.9\% & 37.1\% \\
                            & $(5^\circ, 35^\circ)$ & $(71^\circ, 120^\circ)$ & $(66^\circ, 113^\circ)$ \\
\bottomrule
\end{tabular}
\end{center}
\caption{99\% maximum PTE values (in percentage) for the $S^{XX}(\theta_1, \theta_2)$ estimators found by exploring the $(\theta_1, \theta_2)$ parameter space, in steps of $1^\circ$, for the simulations using the common and inpainting masks. The $(\theta_1, \theta_2)$ values that maximise the PTE can be read below the PTE value. The results have been obtained with $\ell_{\rm max}=$ 32.}
\label{tab:SXX_max_PTE}
\end{table}

\section{The Cold Spot and other large-scale peaks in polarisation}
\label{sec:large_scale_peaks}

The Cold Spot (CS), initially detected in the \wmap\ temperature data \cite{2004VielvaCS,Cruz2005ColdSpot} and subsequently confirmed by the \planck\ mission \cite{P13_XXIII_InS}, is an atypical cold region situated in the southern hemisphere. This phenomenon deviates from the expectations established by the conventional $\Lambda$CDM cosmological model, with an apparent {\it p}-value $\lesssim 0.01$ \cite{P15_XVI_InS}, primarily due to its pronounced large-scale curvature deviation, approaching approximately $4$ standard deviations at a resolution of $R = 5^{\circ}$ \cite{2017JCAPAiramTPeaksAnalysis, 2017AiramLocalProperties}.\footnote{$R$ denotes the width of the Gaussian filter applied in ref.~\cite{2017JCAPAiramTPeaksAnalysis}, where the filter function is given by  $w_{\ell}(R) = \exp\left[-\ell\left(\ell + 1\right)/2R^2\right]$. In this context, $R$ corresponds to the standard deviation of a two-dimensional Gaussian function in Euclidean space.
} The underlying physical cause of this temperature decrement remains unresolved.

\begin{figure}
    \begin{subfigure}{.49\linewidth}
        \centering
        \includegraphics[width=\linewidth]{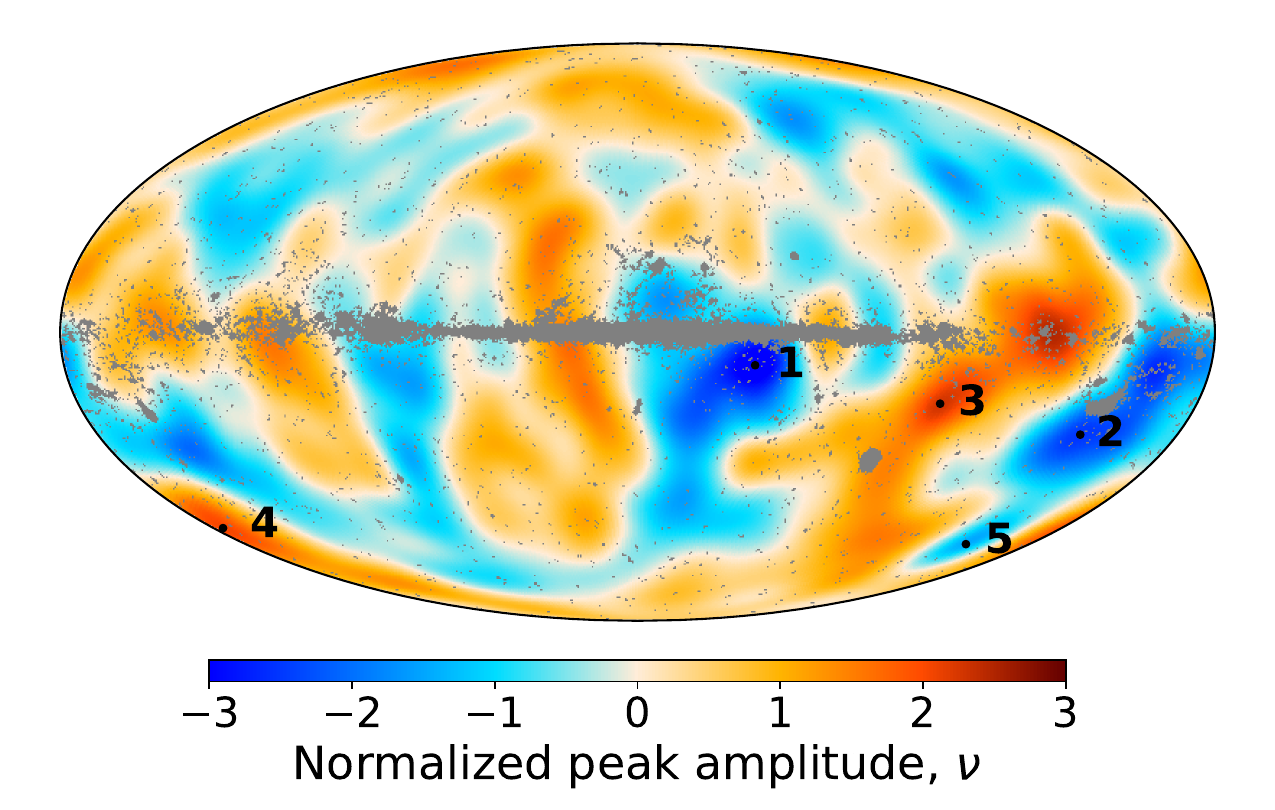}
    \end{subfigure}
    \begin{subfigure}{.49\linewidth}
        \centering
        \includegraphics[width=\linewidth]{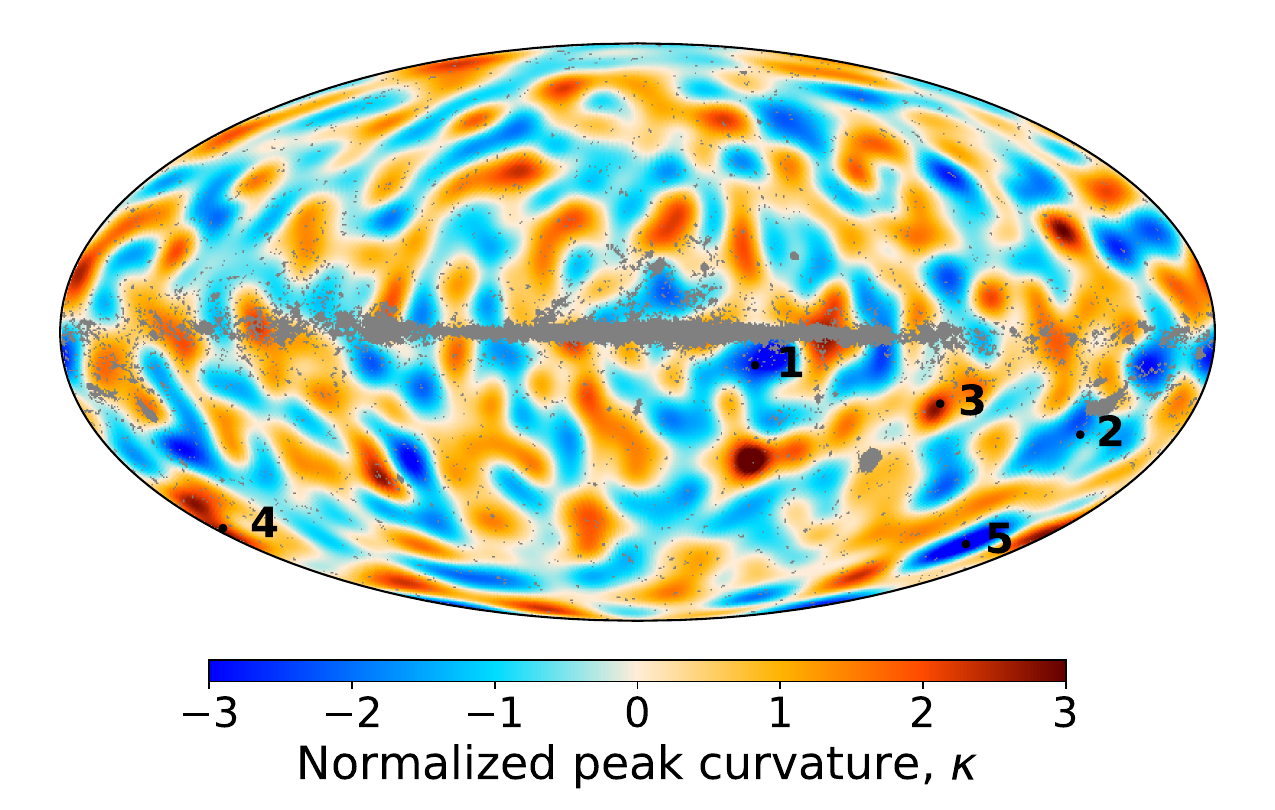}
    \end{subfigure}
    \caption{\smica's normalised $\nu$ (left) and $\kappa$ (right) maps. Maps are smoothed to scale $R=10^{\circ}$ and $R=5^{\circ}$ respectively. The grey area corresponds to the pixels masked by the \Planck\ PR3 inpainting mask. The peaks studied in ref.~\cite{P18_VII_InS} are highlighted as black points.}
    \label{fig:peak_location}
\end{figure}

The \planck\ collaboration additionally conducted an extensive analysis of the $Q$ and $U$ polarisation patterns within the CS and four other large-scale features evident in the \planck\ CMB temperature data \cite{P18_VII_InS}. The additional peaks were selected as the most anomalous structures on large angular scales ($R = 10^{\circ}$) \cite{2017JCAPAiramTPeaksAnalysis}. The central coordinates of the peaks are visually represented in figure~\ref{fig:peak_location}, which displays the normalised amplitude ($\nu$) and curvature ($\kappa$) maps, obtained from the \smica\ data
at two distinct resolutions, namely $R=10^{\circ}$ and $R=5^{\circ}$. Detailed procedures for calculating the normalised $\nu$ and $\kappa$ values can be found in ref.~\cite{2016JCAPAiramPeakTheo}.
However, due to \planck's limited sensitivity to polarisation, no discernible patterns were identified at these locations. 

Nevertheless, if the temperature decrement around the CS were a statistical fluke in the context of the $\Lambda$CDM model, there would be a polarised counterpart signal of predictable amplitude, since $T$ and $E$ are weakly correlated. 
Because \lb\ is expected to detect $E$ modes close to the cosmic-variance limit, here we investigate if such features can be detected in ideal circumstances. 

As indicated by the analysis in appendix~\ref{app:large_scale_peaks_resolution}, improved results can be obtained from data and simulations at a higher resolution than used previously in this paper. Therefore, independent simulations are generated at \nside\ = $512$, as described in section~\ref{subsec:LSP_simulations}. For the constrained realisations, we make use of the inpainted \smica\ map provided in the PR3 release and described in ref.~\cite{P18_IV_CS}. The inpainting mask is shown in figure~\ref{fig:peak_location}. A limitation of working at this resolution is that 
only one inpainted realisation of the \smica\ data is available. Therefore, we focus here on those 
large-scale peaks that are centred far from the Galactic plane, i.e., Peaks 4 and 5 (the CS), where the effects related to foreground residuals and inpainting are expected to be less pronounced. We leave to a future study the effect of the stochasticity of the inpainting process on the results from this analysis. 

The remainder of the section is structured as follows. First, section~\ref{subsec:LSP_simulations} presents the simulations used in this analysis. In section~\ref{subsec:angular}, a comprehensive description is provided regarding the methodology employed for the computation of polarised angular profiles and their associated amplitudes. Finally, section~\ref{subsec:peak_detection} is dedicated to presenting the primary forecasts of \lb's potential to detect the polarised signal in cases where its origin may be attributed to a statistical fluke.

\subsection{Simulations}
\label{subsec:LSP_simulations}

As in the previous analyses, we test the fluke hypothesis using sets of unconstrained and constrained realisations, where the latter correspond to the case where the fluke hypothesis is correct, and the temperature-correlated part of the $E$-mode signal will also be statistically anomalous. These simulations are generated as follows.

 \begin{enumerate}
\item We obtain the temperature spherical harmonics of the inpainted \smica\ temperature map and deconvolve a Gaussian beam of $5^\prime$ from them.
\item Then, we generate constrained $E$-mode spherical harmonic coefficients using 
eq.~(\ref{eq:alm_T_and_E}), where $a^{T}_{\ell m}$ are the temperature spherical harmonic coefficients obtained in the previous step.
\item The $B$-mode spherical harmonic coefficients are generated as Gaussian random realisations from the $C_{\ell}^{BB}$ fiducial power spectrum described in section~\ref{sec:simulations}.
\item $Q$ and $U$ maps are then generated from these spherical harmonics at \nside\ = $512$, convolved with a Gaussian beam of FWHM = $30^\prime$.
\end{enumerate}
We follow this procedure to generate $10\,000$ constrained simulations.

We then generate $1000$ unconstrained realisations from the fiducial power spectra at \nside\ = 512 and convolved with a Gaussian beam of FWHM = $30^\prime$. In this case, although the temperature and $E$ modes are still correlated, a polarised signal may not be detected around the locations of the \smica\ peaks since the temperature realisations might not have peaks there. Of course, polarisation peaks could arise from the $E$-mode contribution that is not correlated with temperature. Note that these simulations differ from the equivalent lower-resolution simulations as the latter have been inpainted.

\subsection{Polarised angular profile}
\label{subsec:angular}

To calculate the polarised angular profile, we use the magnitudes $Q_r$ and $U_r$, given by eq.~(\ref{eq:Qr_Ur}), evaluated at the central coordinates of the respective peak, instead of the traditional Stokes parameters $Q$ and $U$. The magnitude $Q_r$ characterises the radial or tangential polarisation patterns surrounding the observation point, while $U_r$ represents polarisation that is rotated by an angle of $45^{\circ}$ with respect to $Q_r$.  If the peak eccentricity is negligible, as shown in ref.~\cite{2017AiramLocalProperties}, the polarisation field is expected to have rotational symmetry. As a consequence of this symmetry, $Q_r$ and $U_r$ remain independent of the azimuthal angle $\phi$, in contrast to the traditional $Q$ and $U$ parameters.

$Q_r$ exhibits a direct proportionality to the angular power spectrum $C_{\ell}^{TE}$, while $U_r$ is associated with $C_{\ell}^{TB}$. Within the framework of the $\Lambda$CDM model, the $TB$ power spectrum is expected to be zero.\footnote{Notice that recent observations have hinted at the potential detection of isotropic birefringence, which could provide evidence for the existence of parity violations and result in a non-zero $C_{\ell}^{TB}$ \cite{2022_birefringence_patricia}.} Therefore, for this analysis, our primary focus will be exclusively on the parameter $Q_r$.

The angular profile is computed as the average, $\left<Q_r\left(\hat{\theta}\right)\right>$, taken over the azimuthal angles within a specified annulus ($\mathcal{A}(\hat{n}_{\rm{ref}},\theta_{\rm min},\theta_{\rm max})$). Here, $\hat{n}_{\rm{ref}}$ corresponds to the direction aligned with the centre of the annulus, which is also the centre of the corresponding peak. The parameters $\theta_{\rm min}$ and $\theta_{\rm max}$ represent the radial distances that define the inner and outer radii of the annulus, respectively. Furthermore, $\hat{\theta} = \theta_{\rm min} + (\theta_{\rm max} - \theta_{\rm min}) / 2$ is the midpoint of the annular region.

With the angular profile, $\left<Q_r\left(\hat{\theta}\right)\right>$, we can quantify the strength of the polarisation signal associated with the peaks using an amplitude $A$, which is determined by rescaling a model profile. If the profiles follow a Gaussian distribution, the amplitude is also Gaussian and its mean can be calculated as 
\begin{align}
    A = & \dfrac{\sum\limits_{i}\sum\limits_{j} \left<\hat{Q}_r(\hat{\theta}_i)\right> \left[\mymatrix{C}^{-1}\right]_{ij} \left<Q_r(\hat{\theta}_j)\right>}{\sum\limits_{i}\sum\limits_{j} \left<Q_r(\hat{\theta}_i)\right>\left[\mymatrix{C}^{-1}\right]_{ij} \left<Q_r(\hat{\theta}_j)\right> }\, . 
    \label{eq:A}
\end{align}
The associated uncertainty in the amplitude is given by
\begin{equation}
    \sigma_A =  \frac{1}{\sqrt{\sum\limits_{i}\sum\limits_{j} \left<Q_r(\hat{\theta}_i)\right> \left[\mymatrix{C}^{-1}\right]_{ij} \left<Q_r(\hat{\theta}_j)\right>}}\,  ,
    \label{eq:sigma_A}
\end{equation}
where $\left<\hat{Q}_r(\hat{\theta}_i)\right>$ and $\left<Q_r(\hat{\theta}_i)\right>$ are the observed and model angular profiles at the annulus $\mathcal{A}(\hat{n}_{\rm{ref}},\theta_{\rm min, i},\theta_{\rm max,i})$ respectively, and $\mymatrix{C}$ is the covariance matrix encompassing the angular profiles of the various annuli.

To construct a model for each peak's angular profile, $\left<Q_r(\hat{\theta})\right>$, and its associated covariance matrix, $\mymatrix{C}$, we rely on $9000$ constrained simulations. 
It is important to emphasise that the model generated by averaging the profile from simulations does not represent the ensemble-averaged theoretical angular profile as defined in eq.~(\ref{eq:qr0_theo_binned}). Instead, it constitutes a specific realisation of the theoretical profile, in particular, the inpainted \smica\ map realisation. 

Once an angular profile model of the peak is established, we use $1000$ different simulations to calculate the angular profiles of the most extreme peaks. For the constrained realisations, these are identified as peaks $4$ and $5$. However, for the unconstrained simulations, such extrema may not be found at the corresponding locations. Therefore, for each simulation, we identify the most extreme temperature peaks based on their amplitude or curvature at resolutions of $R=10^{\circ}$ and $R=5^{\circ}$, respectively. These new Peaks are then considered equivalent to peaks $4$ and $5$ in the constrained realisations.
The $Q_r$ angular profiles are then calculated at these identified locations. This allows us to obtain the distribution of the amplitude, $A$, by fitting them to the model angular profiles of Peaks $4$ and $5$.

\subsection{Large-scale peak detection forecasts for \lb}
\label{subsec:peak_detection}

\begin{figure}
    \begin{subfigure}{.5\linewidth}
        \centering
        \includegraphics[width=\linewidth]{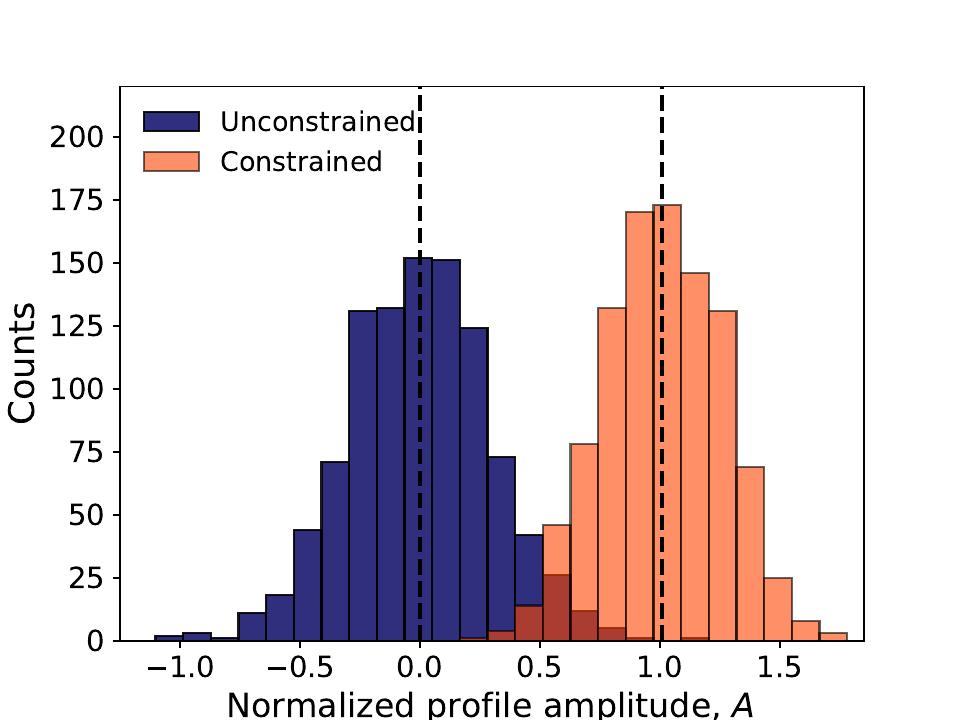}
        \caption{Peak 4.}
        \label{fig:a_histogram_peak4_60_1}
    \end{subfigure}
     \begin{subfigure}{.5\linewidth}
        \centering
        \includegraphics[width=\linewidth]{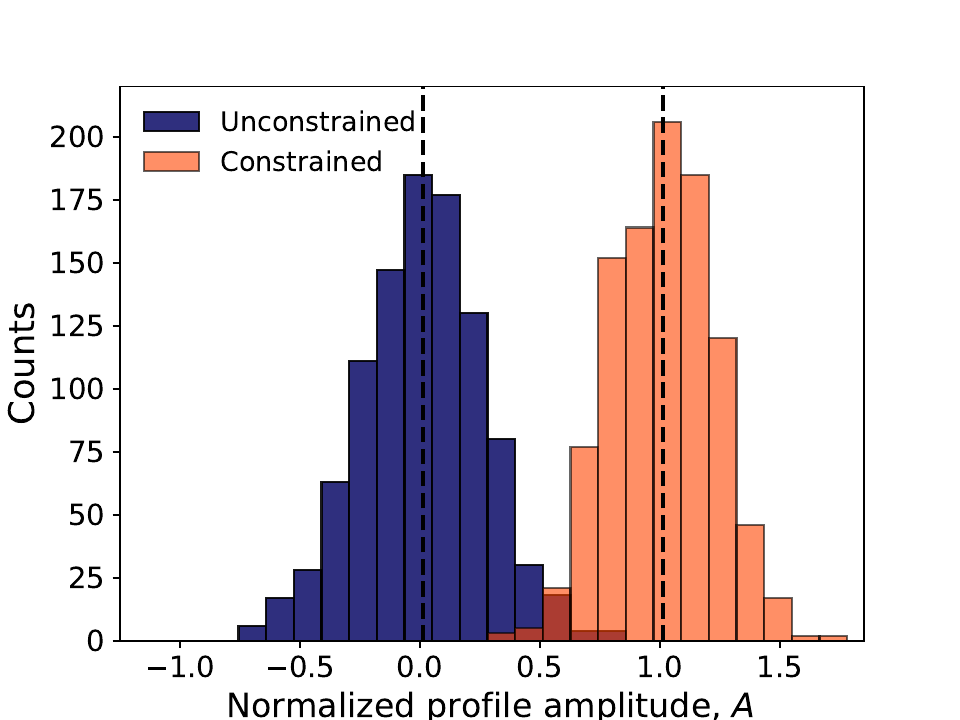}
        \caption{Peak 5 (CS).}
        \label{fig:a_histogram_peak5_60_1}
    \end{subfigure} 
    \caption{Distribution of $A$ obtained from the $1000$ unconstrained (blue) and constrained (orange) simulations, with the overlap shown in red. }
    \label{fig:a_histogram_60_1}
\end{figure}

Figure~\ref{fig:a_histogram_60_1} illustrates the distribution of $A$ from $1000$ constrained and unconstrained realisations for these specific peaks. Additionally, table~\ref{tab:peak_amplitude} presents a summary of the statistics derived from these distributions, including the mean amplitude $\left<A\right>$, the standard deviation $\sigma$ of the distribution, and the signal-to-noise ratio of the peak detection, defined as $\left<A\right>/\sigma$.

\begin{table}
    \centering
    \begin{tabular}{ccccc}
        \toprule
        & \multicolumn{2}{c}{Constrained} & \multicolumn{2}{c}{Unconstrained} \\
        \midrule
        Peak & 4 & 5 & 4 & 5 \\
        \midrule
        $\left<A\right>$ & 1.01 & 1.01 & 0.01 & 0.00 \\
        $\sigma$ & 0.24 & 0.21 & 0.30 & 0.26 \\
        $\left<A\right>/\sigma$ & 4.10 & 4.76 & 0.02 & 0.00 \\
        \bottomrule
    \end{tabular}
    \caption{Summary statistics of the peak amplitude distribution calculated with 1000 simulations of constrained and unconstrained simulations. From top to bottom: the mean $\left<A\right>$, the standard deviation of the distribution $\sigma$, the signal-to-noise ratio of the peak detection defined as $\left<A\right>/\sigma$.}
    \label{tab:peak_amplitude}
\end{table}

The distributions of the peak amplitude are clearly distinguishable, with Peaks $4$ and $5$ being detectable at significance levels exceeding $4\,\sigma$. However, there is a small overlap around $A\simeq 0.5$. 
To assess the statistical significance of an $A$ value belonging to either distribution, we calculate the probabilities of false positives and false negatives for different $A$ thresholds. A false positive occurs when a constrained simulation has a very low $A$ value and is mistakenly assigned to the distribution for the unconstrained simulations. Conversely, a false negative happens when an unconstrained simulation has a high $A$ value and is incorrectly assigned to the distribution for the constrained simulations, represented by the orange distribution in figure~\ref{fig:a_histogram_60_1}. The results of these calculations are shown in table~\ref{tab:hypothesis_test}.

\begin{table}
    \centering
    \begin{threeparttable}
        \begin{tabular}{cccc}
            \toprule
            Peak & $A$ & False Negatives ($\alpha$) & False Positives ($\beta$) \\
            \midrule
            \multirow{3}{*}{4} & 0.451 & 0.01 & 0.062 \\
            & 0.604 & 0.05 & 0.021 \\
            & 0.753$^{a}$ & 0.151 & 0.006 \\
            \midrule
            \multirow{3}{*}{5 (CS)} & 0.534 & 0.01 & 0.019 \\
            & 0.615$^{a}$ & 0.025 & 0.008 \\
            & 0.660 & 0.05 & 0.005 \\
            \bottomrule
        \end{tabular}
        
        \begin{tablenotes}
            \footnotesize
            \item[$a$] Minimum $A$ that could be detected with more than $3\,\sigma_A$.
        \end{tablenotes}
    \end{threeparttable}
    \caption{Peaks' probability of getting a false negative ($\alpha$), or a false positive ($\beta$) for different $A$ values.  }
    \label{tab:hypothesis_test}
\end{table}

In the case of the CS, the probability of not detecting its polarised counterpart ($A/\sigma_A < 3$) if the fluke hypothesis is correct is $0.025$. In other words, if we do not detect the polarised signal, we can reasonably reject the fluke hypothesis with a {\it p}-value of $0.025$. However, in the case of Peak $4$, the non-detection is less conclusive with a {\it p}-value of $0.151$. Conversely, if \lb\ detects the amplitude of a peak at a significance level greater than $3\,\sigma$, we can confidently dismiss an origin different from the fluke hypothesis, with a {\it p}-value of $0.008$ and $0.006$ for Peaks $5$ and $4$, respectively.

In this ideal case, \lb\ possesses the capacity to either rule out the fluke hypothesis or confirm an origin consistent with it. It is essential to note that this analysis will need to be repeated in a more realistic scenario, encompassing residuals after component separation and the inpainting stochasticity effect in future work.

\section{Conclusions}
\label{sec:conclusions}

The presence of large angular scale features in the CMB temperature
anisotropies that appear anomalous in the context of the best-fitting
$\Lambda$CDM model is well established from both \wmap\ and \Planck\
data. Evaluating whether these anomalies, of modest statistical
significance, represent a true challenge to cosmology or are simply
the consequence of a random realisation within the standard
model requires additional, independent, information, since the
temperature measurements on these scales are at the cosmic-variance
limit. Polarisation observations can provide this information, but
current data remain limited by residual systematic artefacts. In this
paper, we consider whether the future CMB satellite, \LB, could
specifically address the fluke hypothesis.

We therefore compare the properties of a number of statistical tests
derived from simulations of the polarised CMB sky where the
temperature-correlated part of the $E$-mode signal is constrained by
current observations to those from unconstrained realisations. In
particular, we consider the \smica\ temperature map from the 2018 \Planck\
analysis as a template for generating this temperature-correlated polarisation anisotropy. However, since
Galactic foreground residuals are likely to remain in the component-separated data, we use an inpainting technique to replace the
high-foreground regions with a Gaussian constrained realisation. These
regions are defined by the \planck\ 2018 common mask, and a dedicated
low resolution inpainting mask derived here. Since the
inpainting approach is stochastic, in practice, 1200 inpainted
\smica\ realisations are used as constraints for one set of $E$-mode
simulations. A second set of 1200 unconstrained realisations for a fiducial
cosmological model provide the basis for comparison. We also
inpaint the corresponding temperature realisations for consistency,
but use only one inpainting realisation per sky.
Note that no foreground residuals after component separation or
systematics are included in the simulations in order to optimise the
possibility of detecting anomalous behaviour. 

In sections 5 through 9 we apply well-known statistical tests
that have previously indicated anomalies in the CMB temperature sky 
to the simulations, inpainted either with the common or inpainting
masks. Ideally, such
tests would be applied to the full sky to provide
an unbiased sample of the CMB signal. However, since the \smica\ map
is statistically unusual, yet inpainted with values from arbitrary
realisations of the standard cosmological model, in almost all cases
the full-sky tests are sensitive to this inpainting process. We have therefore
derived masks for the analysis of the polarised simulations.
In addition, since the transformation to temperature-correlated $E$-mode
signal is non-local and scale-dependent, the masks are determined from
the properties of this component of the polarised sky in
the unconstrained realisations. The analysis of masked data is also
motivated by previous studies that have demonstrated that certain
anomalies become more significant with decreasing sky coverage.

We find that no evidence for or against the fluke hypothesis can be obtained by an
examination of the $E$-mode data alone. 
This implies that the anomalous features observed in temperature
do not significantly impact the properties of the
$E$-mode data. Thus any anomalous statistical properties seen in
polarisation directly provide independent evidence against the standard
cosmological model.
However, for several of the applied
statistical tests, $TE$ (or $TQ$) results show sufficiently distinct
properties between the constrained and unconstrained realisations that
we can reach stronger conclusions about the fluke hypothesis. 

In section~\ref{sec:moments}, we find that if the observed $TE$
covariance measured at $N_\mathrm{side}=64$ with the common mask
applied exceeds $4.62~\mu$K$^2$, then we can reject the fluke
hypothesis at the 99\% confidence level with a probability of $48.7\%$.
In addition, we note that qualitatively comparable sensitivity to the fluke
hypothesis is afforded by various temperature and $E$-mode
permutations of the skewness and kurtosis statistics.

We test for evidence of dipolar modulation in the simulated data in
section~\ref{sec:dipmod_varasym} by fitting dipoles to maps of the
local variance. Although results for the dipole modulation amplitudes do not have
sufficient statistical power to assess the fluke hypothesis,  we find
that the anti-alignment of the dipole directions in temperature and $TE$,
as measured by the separation angle $\alpha_{(T,TE)}$, 
does allow rejection with 13.6\% probability at the 99\% confidence
level for $\cos \alpha_{(T,TE)} < -0.23$.

For completeness, we note that the alignment statistics presented in section~\ref{sec:lowl_alignment} can not be generalised to the $TE$ case, so no conclusions about the fluke hypothesis can be reached.

In section~\ref{sec:harmonic}, we consider statistics constructed from
a harmonic analysis of the data and utilising optimal QML estimates of
the angular power spectra. The point-parity asymmetry can be quantified from the
properties of the even and odd power spectra of the data.
In particular, the quantity  $\delta C^{+}_{TE}$
indicates that the fluke hypothesis can be rejected  at the 99\%
confidence level for the common mask with
$\ell_{max}=25$ at a probability of 51.0\%, and 48.1\% for the
inpainting mask with $\ell_{\rm max}=27$. Then, we test for a lack of correlation on large-angular scales based on the estimator $S^{XY}(\theta_1, \theta_2)$.
For $XY=TQ$ and $\ell_{\rm max}=32$, an optimal statistic is found for the range ($71^\circ,120^\circ$) when applying the extended inpainting mask, yielding a 99\% PTE of 39.9\%. For the case $XY=TE$, a 37.1\% PTE is determined over the range ($66^\circ,113^\circ$). Similar results are found with the application of the common mask.

Finally, section~\ref{sec:large_scale_peaks} considers the radial
polarisation patterns in $Q_r$ (directly proportional to the $TE$
power spectrum) that are associated with two anomalous peaks in the
\smica\ temperature maps, both at high latitude and including the
well-known Cold Spot. If \LB\ detects the amplitudes of these patterns
at a significance larger than $3\,\sigma$, then alternatives to the
fluke hypothesis are strongly disfavoured. Conversely, the
non-detection of these amplitudes at that confidence level would imply
the rejection of the fluke hypothesis with at least a probability of
99.2\%.

A full confirmation of these claims
would require an assessment of more realistic data, which will be
addressed in future work. In due course, tests applied to the \LB\
data itself should also benefit from the use of temperature anisotropy data
improved by the application of component-separation techniques to 
a larger number of observational frequencies. The resulting reduced
levels of residual foreground contamination would then allow 
a larger sky fraction to be considered.

\acknowledgments
This work is supported in Japan by ISAS/JAXA for Pre-Phase A2 studies, by the acceleration program of JAXA research and development directorate, by the World Premier International Research Center Initiative (WPI) of MEXT, by the JSPS Core-to-Core Program of A. Advanced Research Networks, and by JSPS KAKENHI Grant Numbers JP15H05891, JP17H01115, and JP17H01125.
The Canadian contribution is supported by the Canadian Space Agency.
The French \textit{LiteBIRD} phase A contribution is supported by the Centre National d’Etudes Spatiale (CNES), by the Centre National de la Recherche Scientifique (CNRS), and by the Commissariat à l’Energie Atomique (CEA).
The German participation in \textit{LiteBIRD} is supported in part by the Excellence Cluster ORIGINS, which is funded by the Deutsche Forschungsgemeinschaft (DFG, German Research Foundation) under Germany’s Excellence Strategy (Grant No. EXC-2094 -- 390783311).
The Italian \textit{LiteBIRD} phase A contribution is supported by the Italian Space Agency (ASI Grants No. 2020-9-HH.0 and 2016-24-H.1-2018), the National Institute for Nuclear Physics (INFN) and the National Institute for Astrophysics (INAF).
Norwegian participation in \textit{LiteBIRD} is supported by the Research Council of Norway (Grant No. 263011 and 351037) and has received funding from the European Research Council (ERC) under the Horizon 2020 Research and Innovation Programme (Grant agreement No. 772253, 819478, and 101141621).
The Spanish \textit{LiteBIRD} phase A contribution is supported by MCIN/AEI/10.13039/501100011033, project refs. PID2019-110610RB-C21, PID2020-120514GB-I00, PID2022-139223OB-C21, PID2023-150398NB-I00 (funded also by European Union NextGenerationEU/PRTR), and by MCIN/CDTI ICTP20210008 (funded also by EU FEDER funds).
Funds that support contributions from Sweden come from the Swedish National Space Agency (SNSA/Rymdstyrelsen) and the Swedish Research Council (Reg. no. 2019-03959).
The UK \textit{LiteBIRD} contribution is supported by the UK Space Agency under grant reference ST/Y006003/1 -- \lq\lq LiteBIRD UK: A major UK contribution to the LiteBIRD mission-Phase1 (March 25)."  The US contribution is supported by NASA grant no. 80NSSC18K0132.
AG acknowledges support by the MUR PRIN2022 Project \lq\lq BROWSEPOL: Beyond standaRd mOdel With coSmic microwavE background POLarization” -- 2022EJNZ53 financed by the European Union -- Next Generation EU.
Some of the results in this paper have been derived using the \healpix~\cite{gorski2005}, \healpy~\cite{Zonca2019}, \texttt{ECLIPSE}~\cite{Bilbao-Ahedo:2021jhn}, \texttt{numpy}~\cite{numpy}, and \texttt{matplotlib}~\cite{matplotlib} packages.


\bibliographystyle{JHEP}
\bibliography{references}

\appendix

\section{Even-odd asymmetry as a lack of power of the even multipoles}
\label{app:evenodd}

\begin{figure}[t!]
    \centering
  \includegraphics[width=0.75\textwidth]{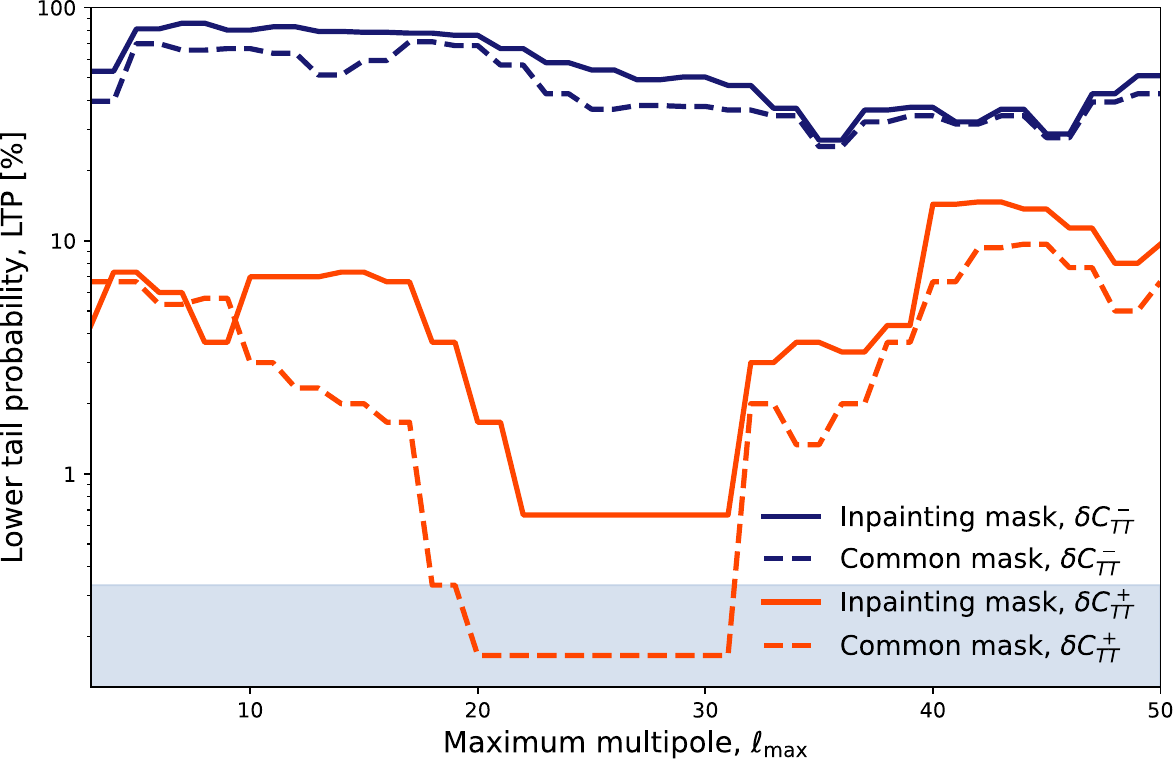}
    \caption{Lower tail probabilities obtained from \texttt{SMICA} \Planck\ data in $TT$. Blue lines are obtained from $\delta C^{-}_{TT}$ and orange lines from $\delta C^{+}_{TT}$. Dashed lines present results derived using the common mask, solid lines for the inpainting mask. The shaded area corresponds to the resolution of the Monte Carlo simulations, due to the number of simulations.}
    \label{fig:percentageCplusCminusTT}
\end{figure}

Here, we consider the estimators $\delta C^{+/-}_{TT}$, as defined in eq.~(\ref{deltaCX}), to measure the even-odd asymmetry in the \Planck\ temperature anisotropy pattern. For each $\ell_{max}$, we compute the LTP of obtaining a value smaller than that observed by \Planck. These LTPs are shown in figure~\ref{fig:percentageCplusCminusTT}, where it is evident that $\delta C^{+}_{TT}$ is anomalously low in the harmonic range $[20,31]$.
In the case of the common mask (see the dashed lines of figure~\ref{fig:percentageCplusCminusTT}), since no simulated values are lower than the observed ones in this range, we set them to half the resolution of the Monte Carlo for plotting purposes. In contrast, the corresponding odd part, $\delta C^{-}_{TT}$, does not exhibit any anomalous behavior. A very similar trend is observed for the inpainting mask results (see the solid lines of figure~\ref{fig:percentageCplusCminusTT}), where, in the same harmonic range, the LTP can be as low as $0.7\%$.
This suggests that the even-odd asymmetry, i.e., the imbalance between even and odd multipoles, can be interpreted as a lack of power in the even multipoles, while odd multipoles remain well within expectations, see, e.g., ref.~\cite{Gruppuso:2017nap}.

If we now construct the estimators $\delta C^{+/-}_{EE}$ considering only the temperature-correlated part of the $E$ mode signal, we observe a behaviour similar to that shown in figure~\ref{fig:percentageCplusCminusTT}. Specifically, for the common mask, the LTP for $\delta C^{+}_{EE}$ is below $1\%$ within the harmonic range $[22,31]$ and drops as low as $0.5\%$ for $\ell_{max} \in [26,31]$, while $\delta C^{-}_{EE}$ does not exhibit any anomaly, i.e., its LTP remains above $63\%$ in the same range.
Similarly, for the inpainting mask, we find that the LTP for $\delta C^{+}_{EE}$ is around $1\%$ in the harmonic range $[22,29]$ and can be as low as $0.7\%$ for $\ell_{max} \in [26,27]$, while the LTP for $\delta C^{-}_{EE}$ remains above $69.8\%$ in the same harmonic range.
This demonstrates that $\delta C^{+/-}_{EE}$ does capture the even-odd parity asymmetry when constructed using the temperature-correlated part of the $E$ mode signal.

\section{Angular profile binning of peaks}
\label{app:large_scale_peaks_resolution}

\begin{figure}
    \begin{minipage}{.48\linewidth}
        \centering
        \includegraphics[width=\linewidth, trim={.4cm 0.2cm 0.35cm 0.2cm}, clip]{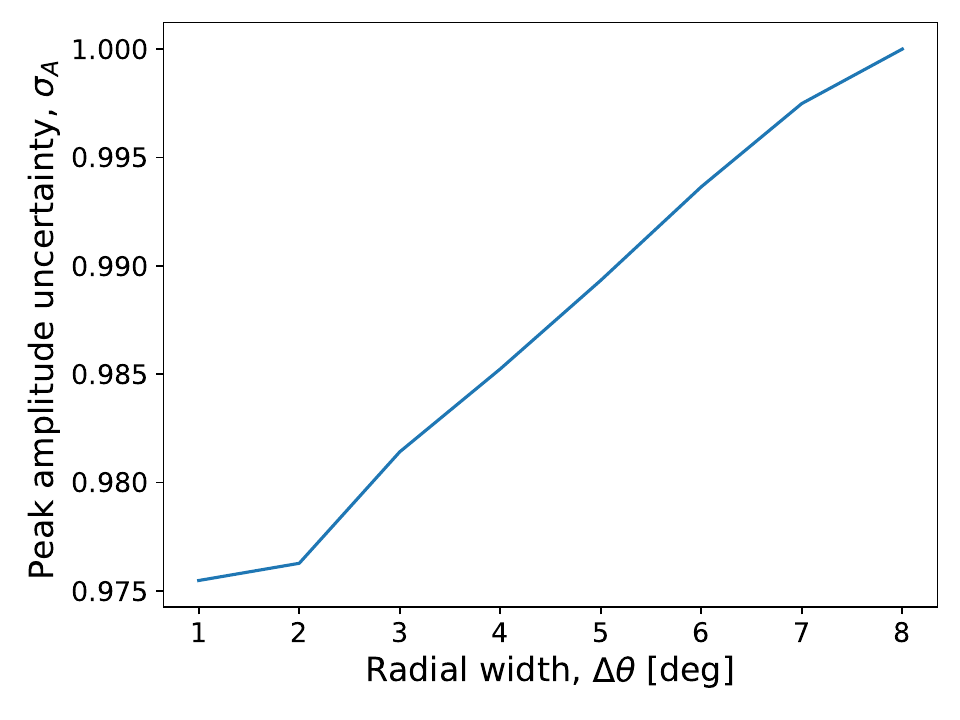}      
        \caption{Normalised ensemble averaged $\sigma_A$ as a function of the uniform difference between the outer and inner radii of the ring scheme considered. }
        \label{fig:sigma_A_binning}
    \end{minipage}
    \hspace{.04\linewidth}
    \begin{minipage}{.48\linewidth}
        \centering
        \includegraphics[width=\linewidth, trim={.3cm 0.2cm 0.35cm 0.2cm}, clip]{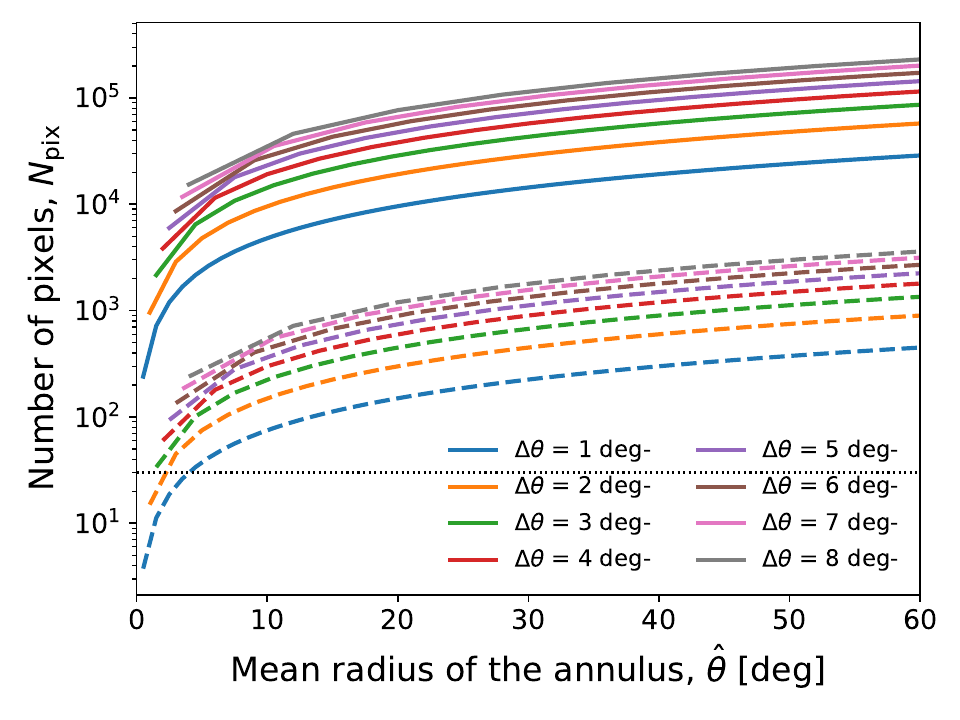}  
        \caption{Number of pixels in a ring as a function of the average angle subtended with respect to the peak's centre for different binning schemes.}
        \label{fig:npix_binning}
    \end{minipage}
\end{figure}


 
We investigate various binning schemes to minimise the uncertainty in the peak amplitude. We consider eight distinct binning schemes, in which the angular profile is determined within annuli that cover the range from $0$ to $60^{\circ}$, each having a uniform width denoted as $\Delta\theta = \theta_{\rm max} - \theta_{\rm min}$. To assess the uncertainty in the peak amplitude, we employ the ensemble-averaged angular profile of eq.~(\ref{eq:qr0_theo_binned}) and its corresponding covariance matrix, eq.~(\ref{eq:qr0_theo_binned}), to calculate $\sigma_A$ in eq.~(\ref{eq:sigma_A}) for the various binning schemes under consideration. We perform this study using the peak characteristics of the CS, i.e., the normalised amplitude and curvature are $\nu = -1.45$ and $\kappa = -4.26$ respectively. The results are illustrated in figure~\ref{fig:sigma_A_binning}.

Figure~\ref{fig:sigma_A_binning} shows that the uncertainty in the amplitude $A$ can be reduced by opting for a binning scheme comprising more smaller rings as opposed to one with fewer larger rings. Nonetheless, the choice of ring size cannot be arbitrarily small, since a sufficient number of pixels within the ring is required for the computation of the angular profile from the maps. Figure~\ref{fig:npix_binning} presents the number of pixels $N_{\rm pix}$ within each annulus as a function of $\hat{\theta}$. The values of $N_{\rm pix}$ are illustrated for the various binning configurations considered, and this information is provided for two different map resolutions, specifically $N_{\rm side}=512$ (solid lines) and $N_{\rm side}=64$ (dashed lines). For the $N_{\rm side}=64$ resolution, it becomes evident that there are insufficient pixels in the rings situated closer to the peak centre for binning schemes characterised by small $\Delta \theta$. Therefore, we choose to proceed with a map resolution of $N_{\rm side} = 512$ for this analysis,  even though only one inpainted map is then available. Consequently, while results for all peaks are presented, our primary focus is directed toward Peaks 4 and 5, which are the peaks furthest from the Galactic plane and hence less impacted by the inpainting realisation.

\section{Theoretical $Q_{r0}$ angular profile}
\label{app:large_scale_peaks}

In this appendix, we present the theoretical equations used to calculate the $Q_{r0}$ angular profile and covariance matrix used in section~\ref{sec:large_scale_peaks}. We follow the equations derived from
refs.~\citep{2016JCAPAiramPeakTheo,2017JCAPAiramTPeaksAnalysis}.

The $Q_{r0}$ theoretical angular profile is given as
    \begin{equation}
    \left<Q_{r0}\left(\theta\right)\right> = -\sum\limits_{\ell=2}^{\infty} \dfrac{2\ell+1}{4\pi} \sqrt{\dfrac{(\ell-2)!}{(\ell+2)!}} \left[b_{\nu} + b_{\kappa}\dfrac{\left(\ell+1\right)!}{\left(\ell-1\right)!} \right]b_{\ell}w_{\ell}(R)C_{\ell}^{TE} P_{\ell}^{2}(\mu(\theta)) \, ,
    \label{eq:qr0_theo}
\end{equation}
where $P^{2}_{\ell}$ is the Legendre polynomial of order $2$ and degree $\ell$, $b_{\ell}$ is the effective window function of the map, $\mu = \cos\theta$ where $\theta$ is the angle subtended from the peak's centre to the direction where the angular profile is calculated, and $w_{\ell}(R)$ is the window function used to emphasise a specific map scale, parametrised by $R$. It is derived from the Fourier coefficients of the stereographic projection of a two-dimensional Euclidean Gaussian distribution:  
\begin{equation}
    w_{\ell}(R) = \exp\left[-\dfrac{\ell\left(\ell + 1\right)}{2R^2}\right] \, .
\end{equation}  
In this context, $R$ represents the standard deviation of a two-dimensional Gaussian function in Euclidean space. $b_{\nu}$ and $b_{\kappa}$ are the bias parameters and are given as
\begin{equation}
    \begin{pmatrix}
        b_{\nu}\sigma_{\nu} \\
        b_{\kappa}\sigma_{\kappa} \\
    \end{pmatrix}
    = \begin{pmatrix}
        1 & \rho\\
        \rho & 1 \\
    \end{pmatrix}^{-1}
    \begin{pmatrix}
        \nu \\
        \kappa
    \end{pmatrix}
    \, ,
\end{equation}
where $\nu$ and $\kappa$ are the amplitude and the curvature of the peak calculated from the temperature map smoothed with $w_{\ell}(R)$ and
\begin{align}
    \sigma^2_{\nu} & = \sum\limits_{\ell} \dfrac{2\ell+1}{4\pi} w_{\ell}(R)^2 C_{\ell}^{TT}  \, ,\\
    \sigma^2_{\kappa} & = \sum\limits_{\ell} \dfrac{2\ell+1}{4\pi} \left(\dfrac{\left(\ell+1\right)!}{\left(\ell-1\right)!}\right)^2 w_{\ell}(R)^2 C_{\ell}^{TT} \, ,\\
    \sigma^2_{\eta} &=  \sum\limits_{\ell} \dfrac{2\ell+1}{4\pi} \left(\dfrac{\left(\ell+1\right)!}{\left(\ell-1\right)!}\right) w_{\ell}(R)^2 C_{\ell}^{TT}  \, ,\\
    \rho &= \dfrac{\sigma^2_{\eta}}{\sigma_{\nu}\sigma_{\kappa}} \,.
\end{align}

Since we are calculating the $Q_{r0}$ in angular rings 
($\mathcal{A}(\hat{n}_{\rm{ref}},\theta_{\mathrm{min}},\theta_{\mathrm{max}})$) 
we need to integrate eq.~(\ref{eq:qr0_theo}) between $\theta_{\rm min}$ and $\theta_{\rm max}$. Reference~\cite{2017AiramLocalProperties} provides analytic formulae of the integral of Legendre polynomials, which we can use to obtain an analytic equation of the angular profile 
\begin{equation}
    \left<Q_{r0}\left(\hat{\theta}\right)\right> = -\sum\limits_{\ell=2}^{\infty} \sqrt{\dfrac{2\ell+1}{4\pi}}  \left[b_{\nu} + b_{\kappa}\dfrac{\left(\ell+1\right)!}{\left(\ell-1\right)!} \right]b_{\ell}w_{\ell}(R)C_{\ell}^{TE} \widehat{\Delta I_{\ell}^{2}}\left(\hat{\mu}(\theta)\right) \, ,
\end{equation}
 where $\hat{\mu}(\theta) = \cos{\hat{\theta}}$ and $\hat{\theta}$ is the average angle in a given angular ring 
 $\mathcal{A}(\hat{n}_{\rm{ref}},\theta_{\rm min},\theta_{\rm max})$. 
 $\widehat{\Delta I_{\ell}^{2}}\left(\hat{\mu}(\theta)\right)$ is given as
\begin{align}
    \dfrac{1}{\mu_{\rm max} - \mu_{\rm min}} \int_{\mu_{\rm min}}^{\mu_{\rm max}} dx P_{\ell}^2\left(x\right)  &= \sqrt{\dfrac{4\pi}{2\ell+1}} \sqrt{\dfrac{\left(\ell+2\right)!}{\left(\ell-2\right)!}} \dfrac{I_{\ell}^2(\mu_{\rm max})-I_{\ell}^2(\mu_{\rm min})}{\mu_{\rm max} - \mu_{\rm min}} \, ,\\
    & = \sqrt{\dfrac{4\pi}{2\ell+1}} \sqrt{\dfrac{\left(\ell+2\right)!}{\left(\ell-2\right)!}} \widehat{\Delta I_{\ell}^{2}}\left(\hat{\mu}\right) \, ,
    \label{eq:qr0_theo_binned}
\end{align}
where,
\begin{align}
    I_{\ell}^2(\mu) &= \dfrac{1}{\sqrt{\left(\ell+2\right)\left(\ell+1\right)\ell\left(\ell-1\right)}}\left[-2I_{\ell}^{0}(\mu) + (\ell+3)\mu \bar{P}_{\ell}(\mu) - \left(\ell+1\right)\sqrt{\dfrac{2\ell+1}{2\ell+3}}\bar{P}_{\ell+1}(\mu)\right] \, ,\\
    I_{\ell}^0(\mu) &= \dfrac{1}{\sqrt{(2\ell+1)(2\ell+3)}}\bar{P}_{\ell+1} - \dfrac{1}{\sqrt{(2\ell+1)(2\ell-1)}}\bar{P}_{\ell-1} \, ,\\
    \bar{P}_{\ell}(\mu) & = \sqrt{\dfrac{2\ell+1}{4\pi}} P_{\ell}(\mu)\, .
\end{align}

As shown in ref.~\cite{2016JCAPAiramPeakTheo}, the covariance matrix has two contributions: an intrinsic contribution that accounts for the inherent correlations between multipolar profiles independently of the existence of a peak; and a peak contribution that modifies the intrinsic covariance to account for the fact that a peak is present in the field. The intrinsic $Q_{r0}$ covariance matrix is
 \begin{equation}
     C^{\rm intr}\left(\theta_{i}, \theta_{j}\right) = C^{\rm intr}_{ij} = \sum\limits_{\ell=2}^{\infty}\dfrac{2\ell+1}{4\pi} \dfrac{\left(\ell-2\right)!}{\left(\ell+2\right)!} b_{\ell}^2 C_{\ell}^{EE} P_{\ell}^{2}\left(\mu\left(\theta_{i}\right)\right) P_{\ell}^{2}\left(\mu\left(\theta_{j}\right)\right) \, ,
 \end{equation}
 where $\theta_i$ is the angle subtended from the peak's centre to the direction where the angular profile is calculated. After integrating over the $\theta$ in the rings $\mathcal{A}_i({\hat{n}}_{\rm{ref}},\theta_{\mathrm{min},i},\theta_{\mathrm{max},i})$, and $\mathcal{A}_j(\hat{n}_{\rm{ref}},\theta_{\mathrm{min},j},\theta_{\mathrm{max},j})$ we obtain
 \begin{equation}
     C^{\rm intr}_{ij} = \sum\limits_{\ell=2} b_{\ell}^2 C_{\ell}^{EE} \widehat{\Delta I_{\ell}^{2}}\left(\widehat{\mu_{i}}\right) \widehat{\Delta I_{\ell'}^{2}}\left(\widehat{\mu_j}\right) \, .
 \end{equation}
The peak contribution to the covariance matrix is

    \begin{equation}
        C^{\rm peak}_{ij} = \sum\limits_{\ell,\ell'}\dfrac{2\ell+1}{4\pi}\dfrac{2\ell'+1}{4\pi} \sqrt{\dfrac{(\ell-2)!}{(\ell+2)!}} \sqrt{\dfrac{(\ell'-2)!}{(\ell'+2)!}} B_{\ell\ell'} b_{\ell} w_{\ell}(R) C_{\ell}^{TE} b_{\ell'} w_{\ell'}(R) C_{\ell'}^{TE} P_{\ell}^{2}\left(\mu_{i}\right) P_{\ell'}^{2}\left(\mu_j\right) \, ,
    \end{equation}
    where,
    \begin{equation}
        B_{\ell\ell'} = -\dfrac{1}{1-\rho^2}\left[\dfrac{1}{\sigma^2_{\nu}}- \dfrac{\rho\left(\ell\left(\ell+1\right) + \ell'(\ell'+1)\right)}{\sigma_{\nu}\sigma_{\kappa}} + \dfrac{\left(\ell\left(\ell+1\right)\right) \left(\ell'\left(\ell'+1\right)\right)}{\sigma_{\kappa}^2}\right] \, .
    \end{equation}
After integrating over the angles in the $i$ and $j$ rings we recover
    \begin{equation}
        C^{\rm peak}_{ij} = \sum\limits_{\ell,\ell'}\sqrt{\dfrac{2\ell+1}{4\pi}}\sqrt{\dfrac{2\ell'+1}{4\pi}}  B_{\ell\ell'} b_{\ell} w_{\ell}(R) C_{\ell}^{TE} b_{\ell'} w_{\ell'}(R) C_{\ell'}^{TE} \widehat{\Delta I_{\ell}^{2}}\left(\widehat{\mu_{i}}\right) \widehat{\Delta I_{\ell'}^{2}}\left(\widehat{\mu_j}\right) \, .
    \end{equation}


\end{document}